# Hybrid phase-change lattice Boltzmann simulation of the bubble nucleation and different boiling regimes of conjugate boiling heat transfer

*Wandong Zhao, Jianhan Liang\*, Mingbo Sun\*, Xiaodong Cai, Peibo Li*

Science and Technology on Scramjet Laboratory, National University of Defense Technology, Changsha 410073, China

**\*Corresponding authors**
*Email: jhleon@vip.sina.com (Jianhan Liang)*
*Email: sunmingbo@nudt.edu.cn (Mingbo Sun)*

**Highlights:**

The entire boiling curve was obtained by phase-change model with conjugate heat transfer.

Hybrid phase-change lattice Boltzmann model was quantitatively validated.

Two treatments of temperature boundary conditions on boiling processes were studied.

Fluctuant heat fluxes and the transverse movement of bubbles were observed.

Thermal responses inside the heater and heat transfer mechanisms were investigated.

**Abstract:**

Pool boiling characteristics in two computational domains with and without considering conjugate heat transfer (CHT) were numerically simulated by an improved hybrid pseudopotential phase-change lattice Boltzmann method (LBM). The effects of constant temperature boundary condition (BC) with nucleate spots and fluctuant temperature on the boiling process were investigated in detail. It was found that for the computational domain without CHT, the treatment of constant temperature BC with nucleate spots is quite easier to produce film boiling than the temperature BC with small fluctuation. However, the results would be the opposite for the case with CHT. The entire boiling curve from the onset of nucleate boiling to fully developed film boiling was presented using the computational domain considering CHT and constant temperature BC with nucleate spots. The simulated critical heat flux showed an excellent agreement with the existing analytical solutions. Hence, the current hybrid phase-change LBM was quantitatively verified. The highly fluctuant heat flux occurred in the CHF and transition boiling as well as the transverse movement of the bubbles had been observed. Furthermore, the thermal responses inside the heater and heat transfer mechanism in different boiling patterns were also comprehensively studied.

# 1. Introduction

   Boiling heat transfer considered as the most effective heat transfer patterns is widely applied in the energy power plates and high technique devices, such as the thermal power plate, nuclear reactor, heat pipes, micro-electric chips, chemical process and so on [1, 2]. Over the past several decades, tremendous efforts have been made to understand the heat transfer mechanism and characteristics associated with the boiling process. The first pioneer's work was conducted by Nukiyaman [3]. They conducted an experimental investigation on the boiling water under a wide range of controlled constant heat flux and presented the first well-known pool boiling heat flux curve related to the degree of wall overheat, which is also called as the boiling curve. Since then, a large number of researchers have carried out studies on the boiling heat transfer via experiment and theoretical model. As a result, numerous experiments were implemented to investigate the boiling curves with different effects such as the properties of the heater, the wettability and structure of the heating surface and wall superheat. Hence, a lot of correction equations and characteristic of microlayer were proposed for different boiling patterns [4, 5], and even the heat transfer mechanism in the nucleate boiling was also studied by using high-speed thermometry method [6, 7]. At the same time, lots of researchers also performed theoretical analysis on the different boiling patterns, but they were still limited in some special boiling patterns such as bubble nucleation [8] critical heat flux (CHF) [9-11] and film boiling [12].

   Due to the limitations of experiments and theoretical analysis, some detailed information including transient temperature variation and local heat flux distribution cannot be obtained accurately. On the other hand, with the improvement of computing performance and numerical model of multiphase flows, numerical simulation has received a lot of attraction and exhibited an excellent potential to study the complicated phenomenon addressing in the boiling process. By using the level-set method, the first work on nucleate boiling heat transfer was conducted by Dhir and co-authors [13, 14]. In addition, Juric and Tryggvason [15] presented an investigation on film boiling by modifying the front tracking method. In the meanwhile, another popular method of the volume of fluid used in multiphase flows has also been widely utilized to simulate the boiling heat transfer [16]. For more detailed results concerning boiling heating transfer based on the traditional numerical methods can be found in Refs. [17, 18]. However, for these methods, owing to the techniques of interface-tracking or interface-capturing, some artificial approaches were employed to trigger boiling heat transfer such as adding tiny bubbles at the initial stage of computation and setting a special waiting cycle time for multiple ebullition cycles. Hence, these methods have a limitation to simulate the entire boiling curve [19].

   In recent years, due to the self-capturing the interface of the liquid-vapor flows [20] and the simplicity of the physical model, the multiphase lattice Boltzmann method (LBM) proposed by Shan and Chen [21] incorporating pseudopotential force has been extensively used for extremely complicated multiphase flows [22]. Furthermore, the single component pseudopotential LBM is extended to handle the liquid-vapor phase change/ phase separation spontaneously by coupling the equation of state (EOS) for real gas in the pseudopotential force. In general, there are two kinds of pseudopotential phase-change LBM models.

   One is double-distribution LBM model, which uses another distribution function (DF) to calculate the target governing energy equation. With respect to this model, it can be further categorized into single-relaxation-time (SRT) model and multiple-relaxation-time (MRT) model for solving the target temperature equation. The improved pseudopotential double SRT phase-change LBM was mainly developed and improved by Gong and Cheng [23, 24].

Based on the improved phase-change LBM, Cheng et al. conducted a lot of investigations on the saturated pool boiling heat transfer [19, 25-28] and saturated vapor condensation [29-31]. The bubble dynamics of growth and departure [24], the effects of wettability of heating surface [24, 32], heater size and degree of superheat on boiling curves and thermal responses inside the heating plate [19] had been comprehensively investigated. It was found that the departure diameter of the single bubble and the cycle time of the bubble release during the boiling process exhibit a power-law relationship with the acceleration of the gravity [24]. With the techniques of conjugate heat transfer for the fluid-solid interface and large ratio of the liquid-solid thermal conductivity, they obtained the entire boiling curve, which present an excellent agreement with the classical Nukiyama's boiling curves [3], and the 2D characteristics of the isotherm inside the heating plate in different boiling regimes were observed [19]. At the same time, Tao et al. also developed the double MRT phase-change LBM model [33, 34]. Considering the conjugate heat transfer, the effects of cavity shapes on different boiling patterns were numerically investigated. However, the entire boiling curves and the thermal responses inside the heating substrate were not presented in their studies.

The other model is hybrid pseudopotential phase-change LBM, which directly solves the target governing energy equation by the finite-difference method proposed by Li et al. [35]. Subsequently, Li et al. [36-39] carried out a lots of research on pool boiling heat transfer based on the newly developed hybrid phase-change model. The effects of the wettability of the heating surface and the hydrophilic-hydrophobic mixed surface on boiling heat transfer were investigated in detail [35, 37]. Besides, the dynamic characteristics of the droplet's evaporation on the heating surface with hybrid wettability were also studied [36, 38]. They demonstrated that decreasing the wettability of the heating surface would result in the decrease of the CHF, and make it easier to yield the filming boiling pattern and produce the onset of nucleate boiling with a low degree of wall superheat [35]. Regarding the droplet's evaporation, the dynamic behaviors of "stick-slip-jump" during the evaporating process were clearly observed by the phase-change model. They also found that this phenomenon was mainly attributed to the Marangoni effect and Young's force [36]. Furthermore, they also revealed that decreasing the wettability of the tops of the solid particles leads to a leftward shift of the boiling curves and a higher heat transfer coefficient after comparing the heat flux of the boiling process occurred in the rough heating plate with different wettability [37]. However, the entire boiling curves and thermal response inside the heating plate were not investigated, and the proposed hybrid phase-change LBM model was also not quantitatively validated.

To date, although so many investigations on pool boiling heat transfer have been conducted, there are still some confuses on modeling a phase-change LBM model, such as the selection of computational domain, the treatments of bubble nucleation and boundary condition (BC) of temperature. At the same time, the initial boiling processes for different boiling regimes were not clearly studied. Therefore, this work is aimed to investigate the effect of computational domain with and without the conjugate heat transfer on boiling heat transfer based on the hybrid pseudopotential phase-change LBM. And the effects of two treatments of temperature BCs with nucleate spots and slightly fluctuant temperature on the boiling process are also discussed in detail. Additionally, the entire boiling curves from the natural convection to fully developed film boiling are simulated by the hybrid phase-change LBM model. It is found that the CHF and the heat flux of film boiling obtained by the current study show an excellent agreement with the theoretical solutions. Therefore, the preciseness of the hybrid phase-change LBM model for boiling heat transfer are quantitatively verified for the first time. Finally, the thermal responses inside the heating plate and heat transfer mechanisms in different boiling patterns are also investigated in current work.

## 2. The hybrid pseudopotential phase-change lattice Boltzmann model

### 2.1. The improved pseudopotential multiple-relaxation-time lattice Boltzmann model for liquid-vapor flows

The pseudopotential LBM model developed by Shan-Chen [21, 40] is widely capitalized on multiphase flows. However, in the original pseudopotential model, the evolution of density DF with SRT operator [41] is employed, resulting in some drawbacks in numerical stability and accuracy [42]. Recently, Li et al. [43] modified the extending forcing term in the moment space, and the flow evolution of density DF with MRT operator is given by [37, 44-47]

$$f_\alpha(x+e_\alpha\delta_t,t+\delta_t) = f_\alpha(x,t) - \overline{\Lambda}_{\alpha\beta}(f_\beta - f_\beta^{eq})\big|_{(x,t)} + \delta_t F'_\alpha\big|_{(x,t)} \tag{1}$$

where $f$ and $f^{eq}$ denote the density DF and the equilibrium density DF respectively. The quantities $\delta_x$ and $\delta_t$ are the lattice space and the time-space, respectively, and both are taken as 1, so $c = \delta_x/\delta_t = 1$ [35]. $e_a$ represents the discrete velocity and $F'_\alpha$ is the forcing term. In current research, the D2Q9 model is employed. Hence, the discrete velocity can be defined as [48, 49]

$$e_i = \begin{cases} (0,0), & i=0 \\ (1,0)c,(0,1)c,(-1,0)c,(0,-1)c & i=1-4 \\ (1,1)c,(-1,1)c,(-1,-1)c,(1,-1)c, & i=5-8 \end{cases} \tag{2}$$

At the same time, $\overline{\Lambda} = M^{-1}\Lambda M$ in Eq. (1) is the collision matrix. $M$ is the orthogonal transfer matrix, and $\Lambda$ is the diagonal relaxation matrix, which is determined by [43, 50]

$$\Lambda = diag(s_0, s_1, s_2, s_3, s_4, s_5, s_6, s_7, s_8) \\ = diag(\tau_\rho^{-1}, \tau_e^{-1}, \tau_\varsigma^{-1}, \tau_j^{-1}, \tau_q^{-1}, \tau_j^{-1}, \tau_q^{-1}, \tau_\upsilon^{-1}, \tau_\upsilon^{-1}) \tag{3}$$

where $s_1 = s_2$, $s_3 = s_5$, $s_7 = s_8$. The flow non-dimensional relaxation time associated with kinematic viscosity ($\upsilon$) has a form as [51]

$$\tau_\upsilon = \frac{1}{s_7} = \upsilon/c_s^2 + 0.5 \tag{4}$$

Note that, the relaxation time has a relationship with local density in the calculation, which has an equation as [51]

$$\tau_\upsilon = \tau_V + \frac{\rho - \rho_V}{\rho_L - \rho_V}(\tau_L - \tau_V) \tag{5}$$

where the quantities $V$ and $L$ represent the gas and liquid phase, respectively.

The DF $f$ and its equilibrium DF $f^{eq}$ can be projected into the moment space $m = M \cdot f$, $m^{eq} = M \cdot f^{eq}$, respectively with the help of transfer matrix $M$, so $m$ and $m^{eq}$ are determined below [52, 53]

$$\underbrace{\begin{bmatrix} m_0(\rho) \\ m_1(e) \\ m_2(\varepsilon) \\ m_3(j_x) \\ m_4(q_x) \\ m_5(j_y) \\ m_6(q_y) \\ m_7(p_{xx}) \\ m_8(p_{xy}) \end{bmatrix}}_{m} = \underbrace{\begin{bmatrix} 1 & 1 & 1 & 1 & 1 & 1 & 1 & 1 & 1 \\ -4 & -1 & -1 & -1 & -1 & 2 & 2 & 2 & 2 \\ 4 & -2 & -2 & -2 & -2 & 1 & 1 & 1 & 1 \\ 0 & 1 & 0 & -1 & 0 & 1 & -1 & -1 & 1 \\ 0 & -2 & 0 & 2 & 0 & 1 & -1 & -1 & 1 \\ 0 & 0 & 1 & 0 & -1 & 1 & 1 & -1 & -1 \\ 0 & 0 & -2 & 0 & 2 & 1 & 1 & -1 & -1 \\ 0 & 1 & -1 & 1 & -1 & 0 & 0 & 0 & 0 \\ 0 & 0 & 0 & 0 & 0 & 1 & -1 & 1 & -1 \end{bmatrix}}_{M} \underbrace{\begin{bmatrix} f_0 \\ f_1 \\ f_2 \\ f_3 \\ f_4 \\ f_5 \\ f_6 \\ f_7 \\ f_8 \end{bmatrix}}_{f}, \quad m^{(eq)} = \rho \begin{bmatrix} 1 \\ -2+3(u_x^2+u_y^2) \\ 1-3(u_x^2+u_y^2) \\ u_x \\ -u_x \\ u_y \\ -u_y \\ u_x^2-u_y^2 \\ u_x u_y \end{bmatrix} \quad (6)$$

where $u_x, u_y$ are the component of velocity and obey $u = \sqrt{u_x^2 + u_y^2}$.

By the aid of the Eqs. (3) and (6), the density evolution in Eq. (1) can be first solved in moment space with a form as [52, 54]

$$\boldsymbol{m}^* = \boldsymbol{m} - \Lambda(\boldsymbol{m} - \boldsymbol{m}^{eq}) + \delta_t (\boldsymbol{I} - \frac{\Lambda}{2})\overline{\boldsymbol{S}} \tag{7}$$

where $\boldsymbol{I}$ is the unit tensor, $\overline{\boldsymbol{S}} = \boldsymbol{M}\boldsymbol{S}$ is the forcing term, and $\boldsymbol{S} = (S_0, S_1, S_2, S_3, S_4, S_5, S_6, S_7, S_8)^T$. After that, the streaming process is implemented in the velocity space with the help of the inverse matrix of $\boldsymbol{M}$ in MRT-LBM model, giving

$$f_i(\boldsymbol{x} + \boldsymbol{e}_i \delta_t, t + \delta_t) = f_i^*(\boldsymbol{x}, t) \tag{8}$$

where $f^* = M^{-1} m^*$. To handle the thermodynamic consistency with a large density ratio, Li et al. [43] modified the source term in the Eq. (7), which is given below

$$\overline{\boldsymbol{S}} = \begin{bmatrix} 0 \\ 6(u_x F_x + u_y F_y) + \dfrac{12\varpi |\boldsymbol{F}_m|^2}{\psi^2 \delta_t (\tau_e - 0.5)} \\ -6(u_x F_x + u_y F_y) + \dfrac{12\varpi |\boldsymbol{F}_m|^2}{\psi^2 \delta_t (\tau_\varsigma - 0.5)} \\ F_x \\ -F_x \\ F_y \\ -F_y \\ 2(u_x F_x - u_y F_y) \\ (u_x F_y - u_y F_x) \end{bmatrix} \tag{9}$$

where $\varpi$ is utilized to keep the numerical stability, and $\boldsymbol{F}_m = (F_{mx}, F_{my})$ is the term of pseudopotential force. Regarding the MRT model, the macroscopic density and velocity are determined by [43]

$$\rho = \sum_i f_i, \quad \rho \boldsymbol{v} = \sum_i \boldsymbol{e}_i f_i + \frac{\delta_t \boldsymbol{F}}{2} \tag{10}$$

where $\mathbf{F}=(F_x, F_y)$ is the total force, which will be further discussed in the next paragraph. In the pseudopotential multiphase LBM, Shan-Chen [40] proposed the pseudopotential force, which is the critical point to simulate the two-phase separation, and the force is defined as

$$\mathbf{F}_m = -G\psi(\mathbf{x},t)\left[\sum_i w(|\mathbf{e}_i|^2)\psi(\mathbf{x}+\mathbf{e}_i,t)\mathbf{e}_i\right] \tag{11}$$

where $G$ is the interaction strength, and $w(|\mathbf{e}_a|^2)$ is the weight factor [50, 54]. In D2Q9 scheme, the weighting coefficients are $w(1)=1/3$ and $w(2)=1/12$. And $\psi$ in Eq. (11) is determined by [55]

$$\psi = \sqrt{\frac{2(P_{EOS}-\rho c_s^2)}{Gc^2}} \tag{12}$$

where $P_{EOS}$ is EOS for the real gas. In current study, the Peng-Robinson (P-R) EOS is employed and defined as

$$P_{EOS} = \frac{\rho RT}{1-b\rho} - \frac{a\varphi(T)\rho^2}{1+2b\rho-b^2\rho^2} \tag{13}$$

where $\varphi(T)=[1+(0.37464+1.54226\omega-0.26992\omega^2)(1-\sqrt{T/T_c})]^2$, and $w=0.344$, $a=0.45724R^2T_c^2/P_c$, $b=0.0778RT_c/P_c$. The parameters $T_c$, $P_c$ calculated by the Eq. (13) are the critical temperature and critical pressure, respectively. According to Li et al.'s work [35], the quantities $a$, $b$ and $R$ are taken as 2/49, 2/21 and 1 respectively. Thus, the critical temperature $T_c$ is set to be 0.1094.

Most recently, Li et al. proposed a new solid-fluid force based on pseudopotential force, which can realize a wide range of contact angle with a small spurious velocity and has a form of [51, 56]

$$\mathbf{F}_{ads} = -G_w\psi(\mathbf{x},t)\left[\sum_i w(|\mathbf{e}_i|^2)\psi(\rho_w)s(\mathbf{x}+\mathbf{e}_i)\mathbf{e}_i\right] \tag{14}$$

where $s(\mathbf{x}+\mathbf{e}_i)$ is a switching scheme, which is set to be 0 and 1 for solid and fluid phase, respectively.

The buoyancy force $\mathbf{F}_g$ should be taken into consideration during the boiling heat transfer process, which is determined by

$$\mathbf{F}_g(\mathbf{x}) = (\rho(\mathbf{x})-\rho_v)\mathbf{g} \tag{15}$$

where $\mathbf{g}=(0,-g)$ and $\rho_v$ are the gravitational acceleration and average density of the entire fluid domain, respectively, which is extensively capitalized on the previous investigations for the LBM phase-change model [19, 27, 57]. As a consequence, the total force in the Eqs. (9) and (10) is equal to be $\mathbf{F} = \mathbf{F}_m + \mathbf{F}_{ads} + \mathbf{F}_g$.

*2.2. The energy equation for heat transfer*

The LBM for the phase change based on the diffusion interface was developed by Zhang and Chen [58], and the governing equation of energy is obtained by

$$\rho\frac{De}{Dt} = -p\nabla\cdot\mathbf{v} + \nabla\cdot(\lambda\nabla T) \tag{16}$$

where $e = C_V T$ is internal energy, $C_V$ is the specific heat capacity at the constant volume, and $\lambda$ is the thermal conductivity [59]. The entropy's local equilibrium energy equation without considering viscous dissipation is governed by

$$\rho T \frac{ds}{dt} = \nabla \cdot (\lambda \nabla T) \tag{17}$$

Based on the general relation of entropy, it can be obtained the following equation

$$ds = \left(\frac{\partial s}{\partial T}\right)_v dT + \left(\frac{\partial s}{\partial v}\right)_T dv \tag{18}$$

According to the Maxwell relationship

$$\left(\frac{\partial s}{\partial v}\right)_T = \left(\frac{\partial p_{EOS}}{\partial T}\right)_v \tag{19}$$

and based on the chain relation and the definition of specific heat capacity, the following relations can be obtained:

$$\left(\frac{\partial s}{\partial T}\right)_v = \frac{(\partial u / \partial T)_v}{(\partial u / \partial s)_v} = \frac{Cv}{T} \tag{20}$$

with the aid of Eqs. (19) and (20), Eq. (21) can be obtained from Eq. (18)

$$ds = \frac{Cv}{T} dT + \left(\frac{\partial p}{\partial t}\right)_v dv = \frac{Cv}{T} dT + \left(\frac{\partial p_{EOS}}{\partial t}\right)_v d\left(\frac{1}{\rho}\right) = \frac{Cv}{T} dT - \frac{1}{\rho^2}\left(\frac{\partial p_{EOS}}{\partial t}\right)_v d\rho \tag{21}$$

Further, Eq. (17) can be rewritten as

$$\rho Cv \frac{dT}{dt} = \frac{T}{\rho}\left(\frac{\partial p_{EOS}}{\partial t}\right)_\rho \frac{d\rho}{dt} + \nabla \cdot (\lambda \nabla T) \tag{22}$$

Using the material derivative $D(\bullet)/Dt = \partial_t(\bullet) + \mathbf{v} \cdot \nabla(\bullet)$, Eq. (22) can be converted to

$$\frac{\partial T}{\partial t} + \mathbf{v} \cdot \nabla T = \frac{1}{\rho Cv} \nabla \cdot (\lambda \nabla T) + \frac{T}{\rho^2 Cv}\left(\frac{\partial p_{EOS}}{\partial t}\right)_\rho \frac{d\rho}{dt} \tag{23}$$

Hence, with the continuity equation, Eq. (24) can be further obtained, which is the target equation of the energy associated with EOS.

$$\frac{\partial T}{\partial t} = -\mathbf{v} \cdot \nabla T + \frac{1}{\rho Cv} \nabla \cdot (\lambda \nabla T) - \frac{T}{\rho Cv}\left(\frac{\partial p_{EOS}}{\partial t}\right)_\rho \nabla \cdot \mathbf{v} \tag{24}$$

By marking the right side of Eq. (24) as $K(T)$ and using the fourth-order Runge-Kutta scheme [60], the time discretization of the governing equation of Eq. (24) can be calculated below

$$T^{t+\delta t} = T^t + \frac{\delta t}{6}(h_1 + 2h_2 + 2h_3 + h_4) \tag{25}$$

where $h_1$, $h_2$, $h_3$ and $h_4$ can be solved by Eq. (26) respectively.

$$h_1 = K(T^t), h_2 = K(T^t + \frac{\delta_t}{2} h_1), h_3 = K(T^t + \frac{\delta_t}{2} h_2), h_4 = K(T^t + \delta_t h_3) \tag{26}$$

Regarding a quantity $\phi$, the spatial gradient and second-order Laplace are given below [61, 62]

$$\partial_i \phi(\boldsymbol{x}) \approx \frac{1}{c_s^2 \delta_t} \left[ \sum_a w_\alpha \phi(\boldsymbol{x} + \boldsymbol{e}_a \delta_t) \boldsymbol{e}_a \right] \tag{27}$$

$$\nabla^2 \phi(\boldsymbol{x}) \approx \frac{2}{c_s^2 \delta_t^2} \left[ \sum_a w_\alpha (\phi(\boldsymbol{x} + \boldsymbol{e}_a \delta_t) - \phi(\boldsymbol{x})) \right] \tag{28}$$

In summary, the multiphase flows are governed by MRT-LBM, whereas the transport of energy equation is calculated by means of the finite-difference method, which is coupled by the $P_{EOS}$ in Eq. (12).

## 3. Computational setup and validation of hybrid phase-change model

### 3.1. Computational setup

In former studies, there are mainly two kinds of computational domains for simulating phase-change process via the LBM. One is the direct heat transfer simulating of pool boiling [24, 35, 39, 62-64], and the other is conjugate heat transfer simulating [19, 26, 27, 65]. In order to compare the difference in the pool boiling process, both two computational domains are considered as presented by Fig. 1. As can be seen from Fig. 1, there are two computational models: one is model A without the solid domain; the other is model B with conjugate heat transfer below the fluid domain. Both two models have a rectangle fluid domain with a grid size of $2\lambda_d \times 1.5\lambda_d$, while an additional grid size of $2\lambda_d \times 0.15\lambda_d$ is employed for the conjugate heat transfer below the fluid domain. Note that the $\lambda_d$ is the Taylor most-dangerous wavelength for the two different densities fluid flow [2, 19], which is used for nondimensionalization of the computational domain and given by

$$\lambda_d = 2\pi \sqrt{\frac{3\sigma}{g(\rho_l - \rho_v)}} \tag{29}$$

which can be rewritten by the characteristic length $l_0$ as

$$\lambda_d = 2\sqrt{3} l_0 \tag{30}$$

where the parameter of $l_0$ is the characteristic length, and it has been widely applied in former studies [19, 27, 31]. It indicates the ratio of surface tension and buoyancy force for the two-phase flow and is defined by

$$l_0 = \sqrt{\frac{\sigma}{g(\rho_l - \rho_v)}} \tag{31}$$

where $\sigma$ is the surface tension, which can be calculated by the test of the Laplace's law. Subsequently, for the convenience of calculation, the characteristic time $t_0$ is introduced, which is defined by Eq. (32).

$$t_0 = \sqrt{l_0 / g} \tag{32}$$

As demonstrated in Fig. 1 (a) and (b), the initial computational domain of the liquid phase and vapor phase are set to be the same as $2\lambda_d \times 0.75\lambda_d$. It should be mentioned that, a grid size of $\lambda_d$ has been proved to be enough resolution to cope with pool boiling [19], therefore, a width of $2\lambda_d$ for the computational domain has a quite high resolution. In present research, the periodic BC is employed in the y-direction of the computational domain for the density DF and temperature field, whereas the no-slip flow is assumed on the top and bottom surface of the fluid domain. Note that, the halfway bounce-back method [66] is applied for the no-slip flow BC for the density DF. In addition, constant temperature $T_{sat}$ and $T_b$ are implemented at the top and bottom wall of the fluid domain for both two models, respectively.

At the initial stage, the stagnant liquid is filled in the bottom of pool with a statured temperature $T_{sat} = 0.86T_c$. The temperature of vapor phase is also kept the same as $T_{sat}$, which means the initial densities of liquid and vapor are set to $\rho_L = 6.5$ and $\rho_V = 0.38$ using Maxwell construction for the EOS, respectively [35]. Meantime, the temperature of solid domain in the model B is also assumed as $T_{sat}$. Following Refs. [35, 37, 58, 67], the physical characteristics of the liquid phase and vapor phase are set to be: special heat $C_{V,L} = C_{V,V} = 6.0$, kinematic viscosity $\upsilon_L = 0.1$, $\upsilon_V = 0.5/3$, while the thermal conductivity $\lambda = \rho C_V \chi$ for the fluid domain is taken as the proportion of the density $\rho$ with $C_V \chi = 0.05$ [35]. The density of solid domain in the Model B is set to be $\rho_S = 3\rho_L$, and the thermal conduction of solid domain is equal to 16.25. Hence, in current research, the dynamic ratio of liquid and vapor phase is $\mu_L / \mu_V \approx 10$, and the thermal conductivity ratio of the liquid and vapor phase is assumed as $\lambda_L / \lambda_V \approx 17$, while the thermal conductivity ratio of solid domain and liquid phase in model B is chosen to be $\lambda_S / \lambda_L \approx 50$. It should be noted that, the previous investigation conducted by Gong and Cheng [19] proved that the thermal conductivity of heating plate has no effect on the quantity of CHF when the thermal conductivity ratios of solid and liquid domain were chosen from $\lambda_S / \lambda_L \approx 30$ to $\lambda_S / \lambda_L \approx 240$. Thereby, the present setup of $\lambda_S / \lambda_L \approx 50$ can eliminate the effect of the thermal conductivity ratio on the CHF. The surface tension is determined by the test of the Laplace's law. Due to the three-phase contact angle ($\theta$) appeared during the pool boiling process, it is set to be $\theta = 55°$ in current study, which is determined by the parameter of the $G_w$ in Eq. (14). According to Ref. [19], the gravity acceleration of $\mathbf{g} = (0, -0.00005)$ is applied in the entire fluid domain. Thus, the Taylor most-dangerous wavelength can be calculated by Eq. (29) and is equal to $\lambda_d \approx 200$. Accordingly, a grid size of $L_x \times L_y = 400 \times 300$ is employed for the fluid domain in present research. It should be pointed out that, in current LBM study, all of parameters are based on lattice unit with the lattice constant $c = \delta_x / \delta_t = 1$, $\delta_x = \delta_t = 1$. Based on Ref. [24], the specific latent heat is obtained by using theoretical method, and it is evaluated to be $h_{LV} = 0.58$ in current hybrid phase-change LBM.

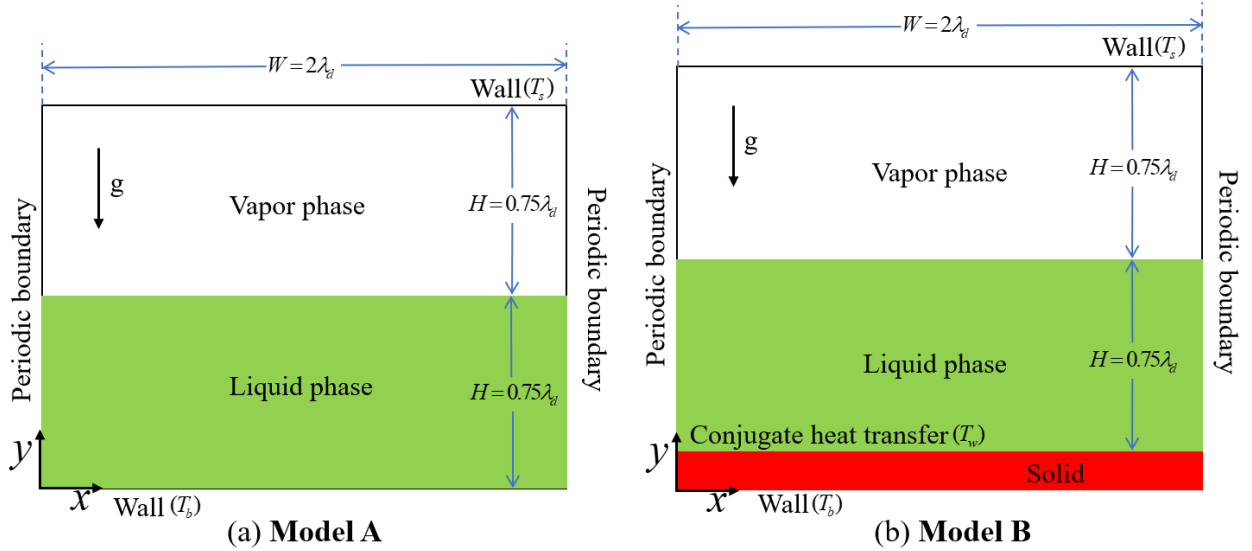

Fig. 1. Schematic of the computational domains ((a) the right column is direct heat transfer without solid domain (b) the left column is conjugate heat transfer domain considering solid domain).

*3.2. Hybrid phase-change LBM model validation*

In former liquid-vapor phase change studies conducted by the hybrid pseudopotential LBM, although the D2 law of droplet's evaporation is used to evaluate the phase-change model, it is slightly inadequate to prove the accuracy of simulating the complicated boiling process. So, following Ref. [19], a quantitative analysis of the film boiling process is also carried out to prove the accuracy of the hybrid phase-change LBM. Without considering the radiation heat transfer during the fully developed film boiling process and assuming that the thermal is transferred to the gas film only via heat conduction, the heat transfer coefficient is theoretically obtained by Berenson [12] for the steady film boiling, which is defined as

$$h_{co} = \frac{Q}{T_w - T_{sat}} = 0.425 \left[ \frac{\lambda_V g \rho_V (\rho_L - \rho_V) h'_{fg}}{\mu_V (T_w - T_{sat})} \right]^{1/4} \left[ \frac{\sigma}{g(\rho_L - \rho_V)} \right]^{-1/8} \quad (33)$$

where $h_{fg}$ is the specific latent heat considering the heating absorb by the thin vapor film, which has a form of $h'_{fg} = h_{fg} + 0.5 C_{V,V}(T_w - T_{sat})$ [2]. At the same time, the space-averaged Nusselt ($Nu$) number is defined as $Nu = h_{co} \lambda_d / \lambda_V$ according to Ref. [19]. Subsequently, with the help of Eqs. (29) and (33), the theoretical equation of the space-averaged $Nu$ number can be derived as follows

$$Nu = 0.425 \times 2\sqrt{3}\pi \left[ \frac{g \rho_V (\rho_L - \rho_V) h'_{fg}}{\lambda_V \mu_V (T_w - T_{sat})} \right]^{1/4} \left[ \frac{\sigma}{g(\rho_L - \rho_V)} \right]^{3/8} \quad (34)$$

In present section, to test the accuracy of hybrid phase-change LBM, the computational model is selected as the same as Ref. [19], which considers the conjugate heat transfer during the boiling process, corresponding to model B as shown in Fig. 1. The constant temperature BC under the bottom of heating substrate is taken as $T_b = 1.27 T_c$, and the other parameters are consistent with Sec. 3.1. In addition, the numerical simulated LBM space-averaged $Nu$ number is defined as [19]

$$Nu_{LBM} = \frac{\lambda_d}{T_w - T_{sat}} \cdot \frac{1}{\lambda_v} \cdot \frac{1}{L_x} \int_0^{Lx} Q(t)\big|_{(x,0)} dx \tag{35}$$

where $Q(t)$ is the local heat flux, which is calculated by $Q(t) = -\lambda(\partial T / \partial y)$. $T_w$ is the space- and time-averaged temperate under the top surface of the solid domain, and it is numerically computed to be $T_b = 1.2660 T_c$ in current simulation. The second order of the finite-difference method is utilized to calculate the one-order thermal gradient, given by

$$\frac{\partial T}{\partial y} = \frac{4T\big|_{(x,1)} - T\big|_{(x,2)} - 3T\big|_{(x,0)}}{2\delta_y} \tag{36}$$

In addition, following the treatment of temperature BC as Ref. [19], in present subsection, a small temperature fluctuation with the equation of $T(x) = T_b + 0.02T_c \sin[2\pi(x - \lambda_d/4)/\lambda_d]$ is also employed to trigger and stimulate the instability of Taylor wave during the time step of $3000\delta_t < t < 15000\delta_t$ (i.e., $5.0 < t^* < 24.98$, in which $t^* = t/t_0$), whereas a constant temperature $T_b$ is used for the next boiling process. It is worth mentioning that, the time- and space-averaged temperature of $T_w$ and heat flux are obtained after $t > 25000\delta_t$ (i.e., $t^* > 41.63$) for the sake of eliminating the influence of fluctuant temperature during the initial time stage. It should be pointed out that the effects of different treatments of temperature BC on the different computational models will be investigated in the next subsections.

The comparison of the transient variations of the simulated space-averaged $Nu$ number with the theoretical result [12] in film boiling is presented by Fig. 2. The space- and time-averaged $Nu$ number obtained from the theoretical result is equal to 51.28. The space- and time-averaged $Nu$ number calculated by the improved hybrid phase-change model equals to 54.01. Accordingly, there is a relative derivation of 5.30%, indicating that the numerical result obtained by the pseudopotential phase-change model agrees well with the Berenson's analytical solution of Eq. (34). In the meanwhile, as shown in Fig. 2, there is a high fluctuation of the $Nu$ number during the simulated film boiling process. This is attributed to the growing up and departure of bubble from the superheat vapor film. In order to clearly demonstrate this phenomenon, the temporal variation of film boiling is given by Fig. 3. One can observe from Fig. 3 that, due to a lot of thin vapor film yielding in the top surface of the heater, it prevents the thermal transfer across the thin vapor film, resulting in a low local heat flux as depicted in Fig. 2 during the time of $33.30 < t^* < 49.96$. On the contrary, with the growing up and subsequent departure of the bubble as shown in Fig .3(d), it enhances the heat transfer ratio and the space-averaged $Nu$ number reaches a maximum state during the time of $58 < t^* < 60$. These results also agree well with the predecessor's literature [19]. In summary, it could be concluded that the current hybrid phase-change LBM is quantitatively verified.

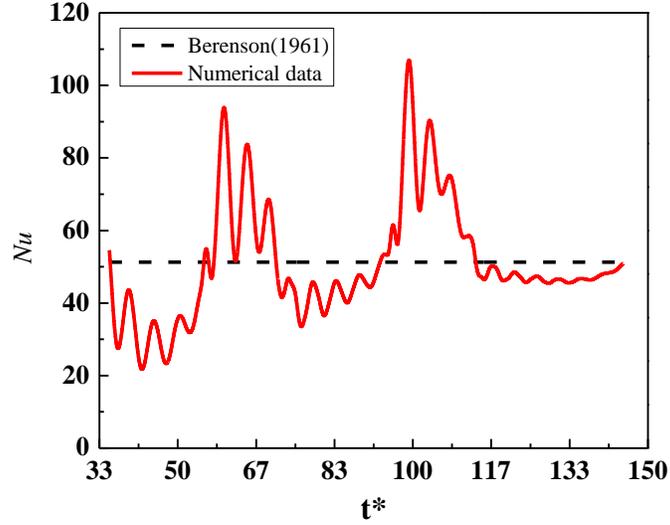

Fig. 2. Comparison of the transient parameter of the space-averaged $Nu$ number by hybrid phase-change LBM with the analytical result [12] in film boiling for wall superheat $T_b = 1.27 T_c$.

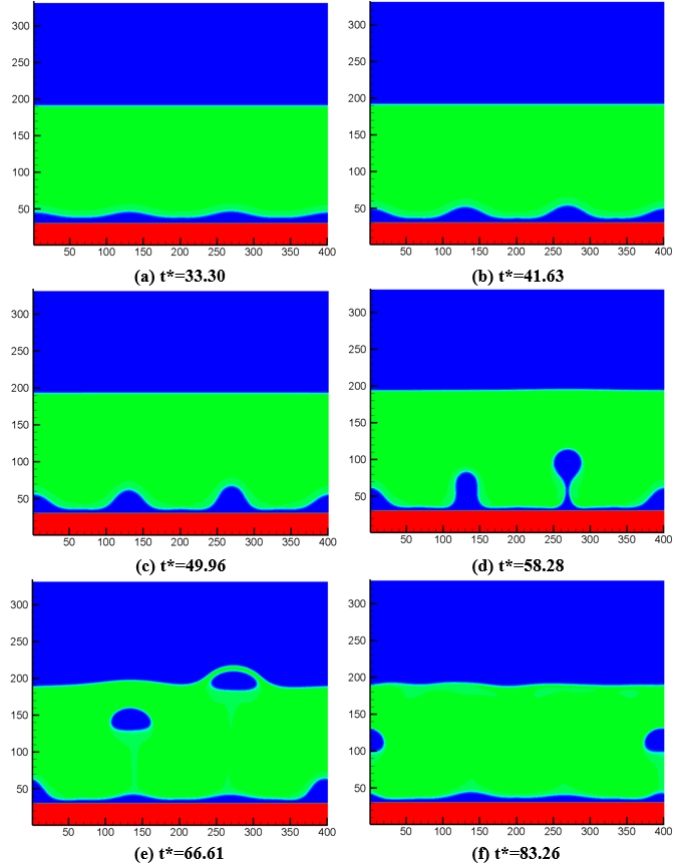

Fig. 3. Temporal evolution of the film boiling with a wall superheat $T_b = 1.27 T_c$ for a conjugate heat transfer.

## 4. Results and discussion

In the next, the validated hybrid phase-change LBM will be employed to simulate the boiling heat transfer with and without the conjugate solid domain. The different treatments of temperature BC on the two computational domains will be investigated firstly. And then the stimulated entire boiling curves and different boiling patterns will be extensively discussed. Finally, the thermal distribution and local heat flux of different boiling regimes in the heating plates will be studied as well.

*4.1. Pool boiling without conjugate heat transfer*

Here, we first investigate the saturated pool boiling in the model A without heat conduction inside the heating plate as presented by Fig. 1. It is also indicated that the thermal is directly specified at the bottom surface of the fluid domain. It is generally acknowledged that the bubble nucleation is the critical point to produce the boiling heat transfer. As a matter of fact, the bubble nucleation cannot be directly produced in a flat and smooth heating surface without some particular treatment [57]. Therefore, in previous literature, the temperature fluctuation is extensively capitalized on the bottom of heating wall [19, 35] to trigger the bubble nucleation. Meanwhile, no-uniform structures including the roughness of the heating substrate [28, 34, 37, 68] and the artificial heterogeneous wall with different wettability [31] are also added in the heating wall to promote the formation of bubble nucleation. Therefore, in current study, two different treatments of the temperature BC are applied in model A. One is set up a small temperature fluctuation during the initial time step as described in Sec. 3.2, and then a uniform temperature is specified in the bottom of fluid domain. It is worth emphasizing that the time sequence of temperature BC is also consistent with Sec. 3.2. The other is used a constant temperature BC ($T_b$) on the bottom surface during the whole boiling process, and three nucleate spots are implemented in the bottom surface to imitate the bubble nucleation. The three nucleate spots with hydrophobic property will result in a different wettability to yield non-uniform interaction force between the fluid and solid. It is implemented by changing the given value of $G_w$ in Eq. (14). The three spots are located in the $x$=66, 200, and 356, respectively, and the contact angles of the three spots are taken as $\theta$=105°, $\theta$=115° and $\theta$=125°, respectively, so as to produce different interactions between the fluid and solid domain. Note that the schematic of the spot distribution will be illustrated in the section 4.2.

After using the two treatments of temperature BC, the time evolutions of pool boiling heat transfer of model A with different wall superheats are studied in detail. Firstly, comparison of the separated shape of bubble regime in the nucleate boiling of two treatments in the Model A with a low wall superheats ($Ja=0.1924$) is given by Fig. 4. Note that the wall superheat is nondimensionalized with Jacob number ($Ja$), which is also used in prior literature [19, 27], defined as

$$Ja = \frac{C_V (T_b - T_{sat})}{h_{fg}} \tag{37}$$

The upper row in Fig. 4 is the BC treatment with fluctuating temperature, and the lower row is the constant temperature BC with three nucleate spots. As shown in Fig. 4(a) and (i), three bubble nucleation successfully appears in both of two BCs, which demonstrates that both two treatments can achieve the formation of bubble nucleation. At the time of $t^*$=24.98, two more bubbles are growing up and close to the $x$=100 and $x$=300 as shown in Fig. 4(b), whereas the number of bubbles in Fig. 4(ii) is still kept the same as the initial snapshot. This result is attributed to an extra small temperature fluctuation applied in upper row of Fig. 4, which leads to more thermal energy absorbed by fluid domain. Subsequently, the adjacent bubbles are merging together at the $x=100$ and $x=300$ as presented in Fig. 4(c), while regarding the second treatment of Fig. 4(iii), a lot of separated bubbles grow up in the heating wall. Finally, owing to the effect of buoyant force, several bubbles overcome the interaction force from the wetting wall and detach from the heating surface as presented in Fig 4(iv) and (d), and several vapor residual are clearly observed in the wake of the departing bubbles. In summary, both two treatments of temperature BC can realize the bubble nucleation, but the frequency of bubble generation in the computational domain with fluctuant temperature BC is higher than the case with constant temperature BC owing to slightly high temperature inside the heating wall.

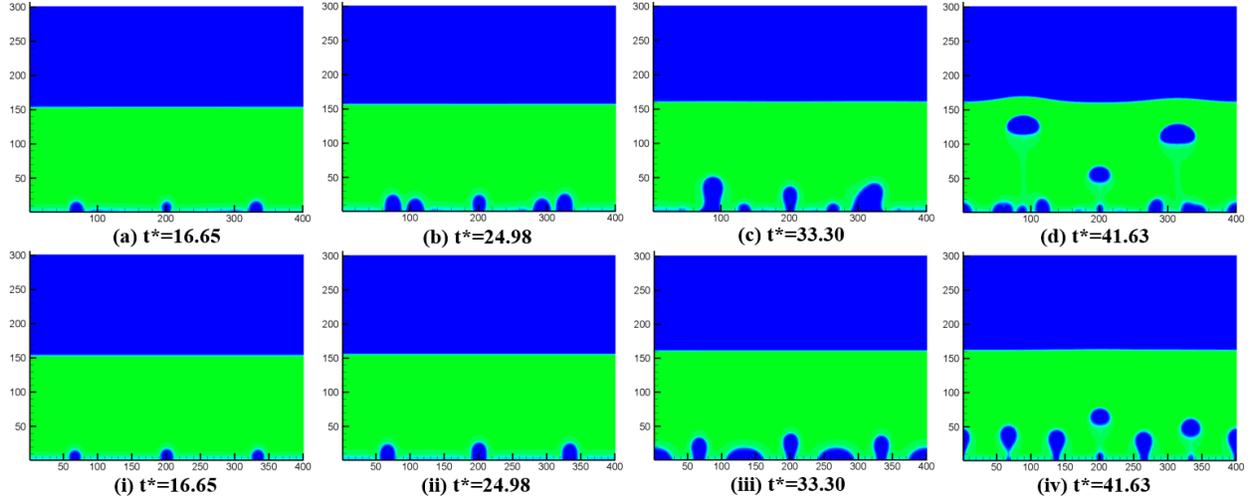

Fig. 4. Comparison of the nucleate boiling regime with a low wall superheat ($Ja = 0.1924$) for different boundary treatments at the same initial dimensionless time (the upper row is fluctuant temperature BC and the lower row is constant temperature BC with nucleate spots).

Next, the boiling process in model A with a higher wall overheat ($Ja = 0.2603$) in both two temperature BCs is studied. Following Fig. 4, the boiling process of model A with two BCs are illustrated in Fig. 5. One can observe from the figure that, with the increasing of wall superheat, the thin vapor film is formed in Fig. 5(i) in the region of $100 < x < 400$, while several isolated bubbles appear in Fig. 5(a). It indicates that the constant temperature BC is quite easier to produce film boiling and it would yield a shorter transition boiling regime. When $t^*=24.98$, the bubble is detaching from the superheat wall as shown in Fig. 5(b), and the thin vapor film is gradually gathering together owing to the influence of surface tension as depicted in Fig. 5(ii). At the next time of $t^*=33.30$, due to the unstable film boiling as shown in Fig. 5(iii), several bubbles are separated in the region of $150 < x < 350$. And then, two bubbles have detached from the heating surface. At the same time, one of bubbles has risen to the top surface of liquid phase as shown in Fig. 5(c). Finally, the transverse movement of bubbles is observed in Fig. 5(d). This phenomenon is also observed in previous research experimentally [69, 70].

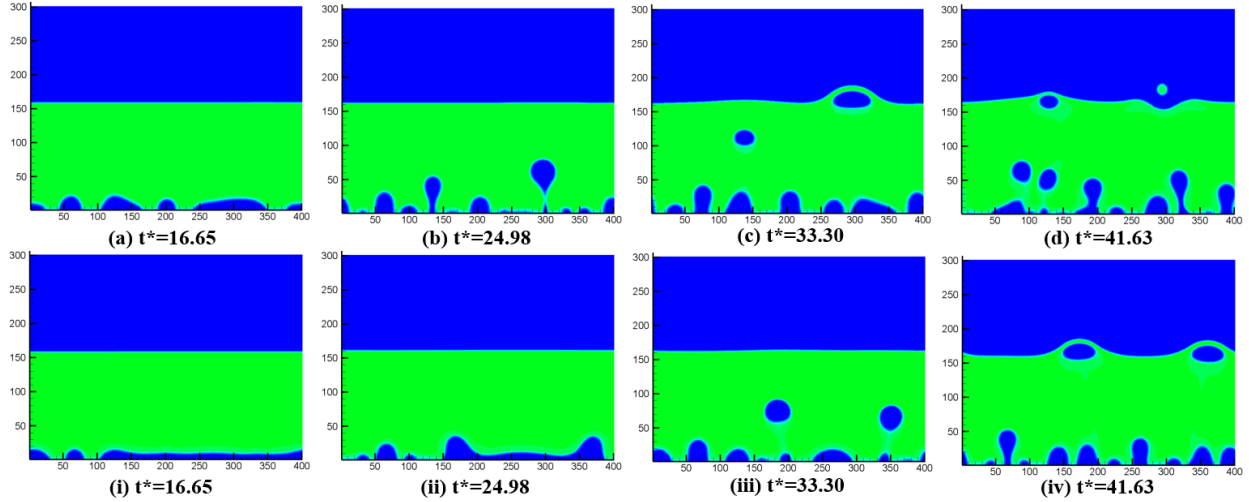

Fig. 5. Boiling processes for the fluctuant temperature BC (the upper row) and constant temperature BC with nucleate spots (the lower row) with a wall overheat ($Ja = 0.2603$) at the same time during the initial stage.

The wall superheat is further augmented to $Ja = 0.3056$. The time history of boiling processes using two different treatments of temperature BC for the model A with $Ja = 0.3056$ are presented in Fig. 6. As shown, the film boiling regime has taken place in the Model A with a constant temperature BC. On the contrary, the boiling regime occurred in model A with a fluctuant temperature BC is still in nucleate boiling. Such a feature is mainly attributed to that the stable film vapor will be formed quickly when the heating surface has a high and uniform constant temperature. In the meanwhile, the present physical model without the solid domain does not the real pool boiling. With respect to model A having a fluctuant temperature, strong interaction between the adjacent bubbles and the merge of closed bubbles on the heating surface appear in Fig. 6 (b-d). This is also demonstrated the last conclusion that a constant temperature BC in the computational domain of model A is beneficial to produce a film boiling compared to the fluctuant temperature BC.

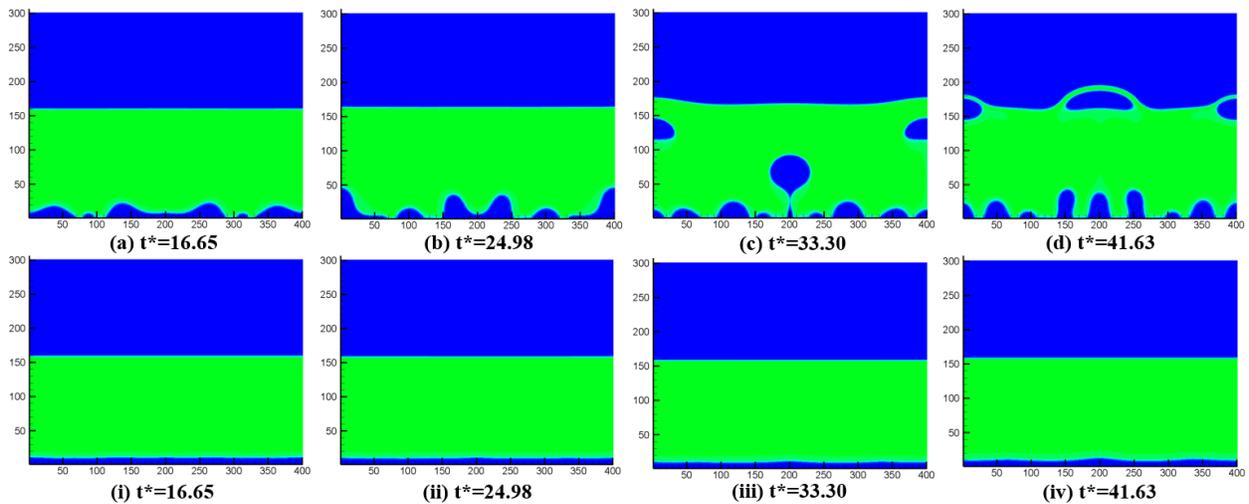

Fig. 6. Temporal evolutions of the boiling characteristic with a high wall superheat ($Ja = 0.3056$) for different boundary treatments at the same initial dimensionless time (the upper row is fluctuant temperature BC and the lower row is constant temperature BC with nucleate spots).

*4.2. Pool boiling with conjugate heat transfer*

In this section, the boiling heat transfer with the solid domain to carry out the conjugate heat transfer is investigated. As above mentioned, the two treatments of the temperature BC are still imposed on the heating plate. The distribution of the three nucleate spots is illustrated in Fig. 7. This structure can produce a different interaction between the solid and fluid when the nucleate bubble is formed. The second BC of the fluctuant temperature during the initial time is kept the same as Sec.4.1.

The transient evolutions of the growing of nucleate bubbles of two treatments for temperature BC with a slight low wall overheat ( $Ja = 0.1924$ ) are presented by Fig. 8. Note that the result of upper row is the case of boiling heat transfer with the fluctuant temperature BC, and the lower row is the BC with a controlled constant temperature and nucleate spots. It can be seen that there is almost no formation of nucleate bubble at the initial time of $t^*$=16.65 for both two treatments as shown in Fig. 8(a) and (i). But at the time of $t^*$=24.98, three nucleate bubbles appear on the Fig. 8(ii), whereas there is still no nucleate bubble formation in the case of fluctuant temperature BC. These behaviors, in the same wall superheat, are different from boiling characteristics without conjugate heat transfer that the formation of the nucleate bubble almost appears at the same initial time as shown in Fig. 4(a) and (i). The first reason is that the initial thermal energy is absorbed by the heater rather than directly absorbed by the fluid. And the second reason is attributed to the nucleate spot which is much easier to form the nucleate bubble due to the unbalance interactive force between the solid and fluid domain. When $t^*$=33.30, two nucleate bubbles are generated on the heating surface for the first case as shown in Fig. 8(c), and nucleate bubbles are further growing up for the second case as depicted by Fig. 8(iii). Finally, as a result of a lot of thermal energy absorbed by the fluid domain, numerous nucleate bubbles appear on the heating surface as shown in Fig. 8(d) and (iv).

Next, the wall superheat is further augmented to $Ja = 0.2603$. The comparison of boiling process for two treatments of temperature BC at the same dimensionless time is given by Fig. 9. It is clearly observed that, at $t^*$=16.65, the nucleate bubble is growing up for the second case with a constant temperature, while there is no nucleate bubble in the case with fluctuant temperature as shown in Fig. 9(a). At the next time, a thin vapor film dramatically generates in the top surface of heating plate as shown in Fig. 9(b). However, At the same time, the isolated nucleate bubble is growing up as shown in Fig. 9(ii), and lots of new small nucleate bubble generate as shown in Fig. 9(iii). Some of bubbles depart from the heating surface due to the effects of gravity and surface tension. On the contrary, because of the generation of vapor film and unsteady of Taylor wave for the case with fluctuant temperature BC, two bubbles are formed, and the formed bubbles depart from the heating surface leading to the breakup of thin vapor film. Generally, at the initial time, film boiling heat transfer has formed for the BC with the fluctuant temperature. But, due to the unsteady film boiling, the thin film is breakup, leading to a transition boiling. However, the nucleate boiling heat transfer is occupied the entire initial time for the case with constant temperature and nucleate spots. Therefore, it can be concluded that the treatment of temperature BC with a fluctuant temperature is much easier to yield film boiling heat transfer compared to the treatment of temperature BC with constant temperature and nucleate spots.

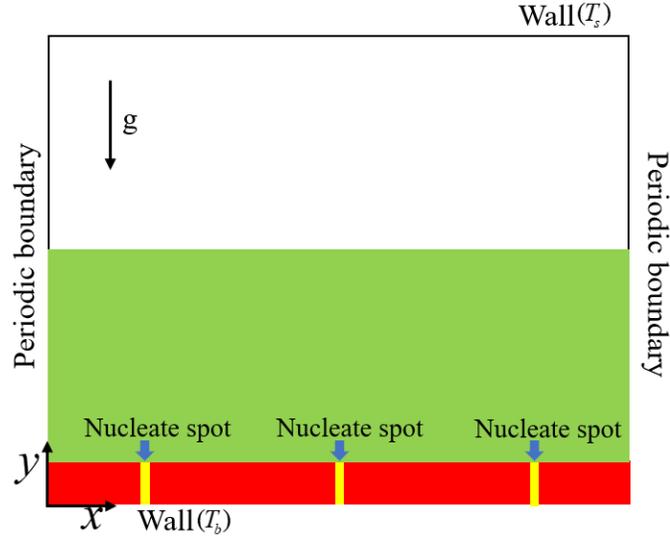

Fig. 7. Schematic of the distributions of the nucleate spot treatment in the heating plate.

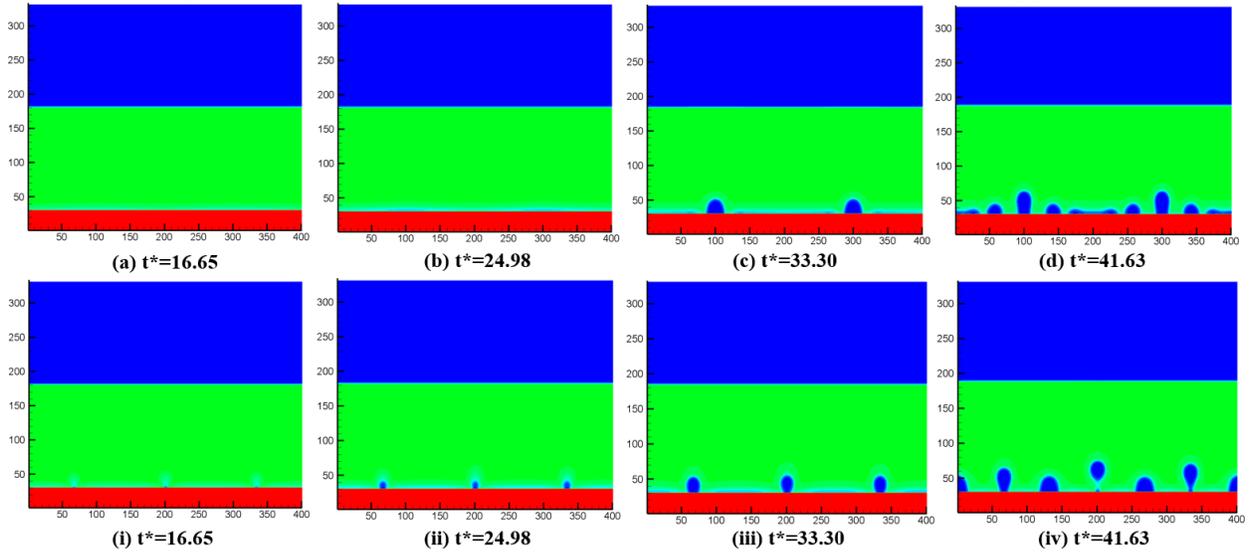

Fig. 8. Comparison of the nucleate boiling with a low wall overheat ($Ja = 0.1924$) for different boundary treatments at the same initial dimensionless time (the first row is fluctuant temperature BC and the second row is constant temperature BC with nucleate spots).

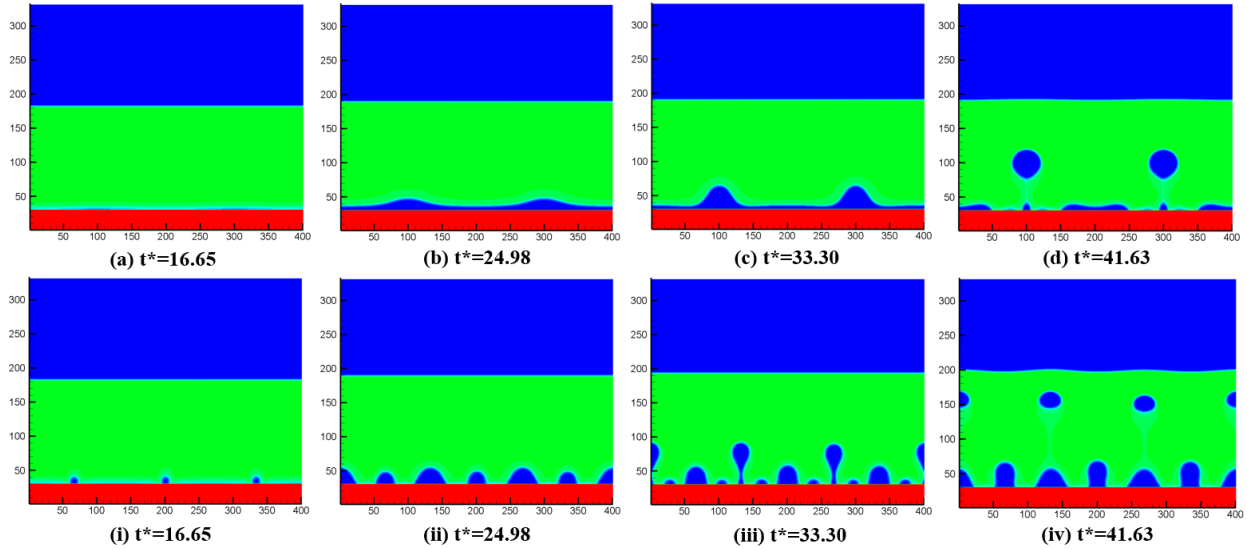

Fig. 9. Boiling characteristics with a middle wall superheat ($Ja = 0.2603$) for different boundary treatments at the same initial dimensionless time (the upper row is fluctuant temperature BC and the lower row is constant temperature BC with nucleate spots).

In order to verify the conclusion, the wall superheat is further increased to $Ja = 0.3508$. The time evolution of the boiling process at the initial dimensionless time for both two treatments of temperature BC is illustrated in Fig. 10. One can observe that the thin vapor film covers all the heating surface during the time of $16.65 < t^* < 41.63$ as shown in Fig. 10(a-d), which is confirmed as film boiling heat transfer. However, for the second case, part of heating surface is covered by the thin vapor film, and part of heating surface is the nucleate bubbles, which is confirmed as the regime of transition boiling heat transfer. These outcomes of boiling process further demonstrate the above conclusion that the treatment of fluctuant temperature is beneficial to produce film boiling. Therefore, the boing heat transfer in this case has a short transition boiling regime compared to the second BC treatment with constant temperature and nucleate spots.

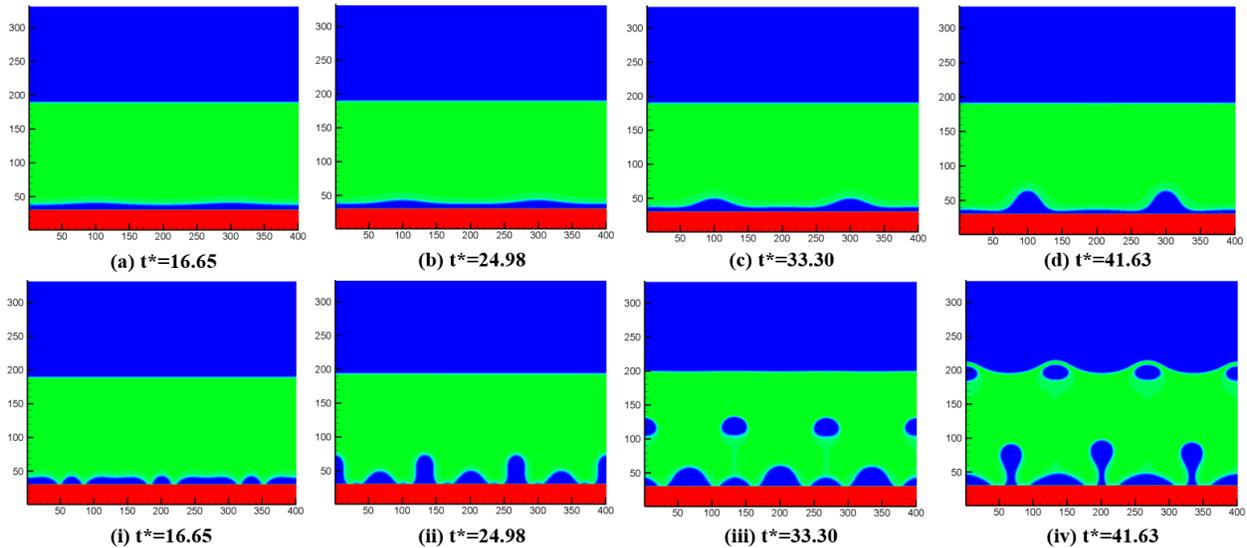

Fig. 10. Snapshots of the boiling regime with a high wall superheat ($Ja = 0.3508$) for different boundary treatments at the same initial dimensionless time (the upper row is fluctuant temperature BC and the lower row is constant temperature BC with nucleate spots).

*4.3. The entire boiling curve and boiling regimes*

4.3.1. The entire boiling curve

In this section, different wall overheats in the bottom surface of the heater for the model B with conjugate heat transfer and the treatment of nucleate spots as mentioned in the Sec. 4.2 are employed to simulate the different boiling regime. The entire boiling curve is obtained by the validated hybrid pseudopotential phase-change LBM model as shown in Fig. 11 under the condition of the thermal conductivity ratio of $\lambda_s / \lambda_L = 50$, $\lambda_L / \lambda_V = 17$ and the wall contact angle of $\theta = 55°$. Note that, in this figure, the wall superheat is defined by Eq. (37) and the $T_b$ is the temperature of the bottom surface of the heater. $Q''$ is the space- and time-averaged dimensionless heat flux in the bottom surface of heating plate, and it has a form of

$$Q'' = \left[ \int_{t_a}^{t_b} Q'(t) dt \right] / (t_b - t_a) \tag{38}$$

where $Q'(t)$ is the dimensionless space-averaged boiling heat flux considering the reference heat flux, which is given below

$$Q'(t) = \frac{Q(t) l_0}{\mu_l h_{fg}} = -\frac{\lambda_s l_0}{\mu_l h_{fg}} \frac{1}{L_x} \int_0^{L_x} \left. \frac{\partial T}{\partial y} \right|_{(x,0)} dx \tag{39}$$

It should be noted that the thermal gradient is obtained by Eq. (36).

As presented by Fig. 11, the boiling curve obtained by the hybrid pseudopotential phase-change LBM model exhibits the similarly general characteristic as the classical Nukiyama boiling curve under the condition of given constant value of wall superheat [3]. When the wall has a low wall superheat, it only occurs the natural convection, and the variation of the heat flux almost has a linear relationship with the wall overheat. With the increasing of wall superheat, the liquid-vapor phase change takes place, and the heat flux increases dramatically as shown in point A in Fig. 11, which is also called as the onset of nucleate boiling (ONB). The heat flux continuously increases as the wall superheat is increasing until it reaches the CHF as highlighted in point C in Fig. 11. And then, the boiling heat flux is gradually decreasing with the increasing of wall superheat owing to the transition boiling regime until it reaches the initial stage of film boiling process. Finally, the boiling heat flux increases again when the wall overheat is further augmented during the regime of fully developed film boiling.

In current section, the numerical simulated CHF is also compared with the former research of the CHF model developed by Zuber [10] and Kandlikar [9], which is defined as follow:

$$Q_{CHF} = K \rho_V h_{fg} \left[ \frac{\sigma g (\rho_L - \rho_V)}{\rho_V^2} \right]^{1/4} \tag{40}$$

where $K$ is the constant parameter and equals from 0.12 to 0.157 for Zuber's model, and K could be related to contact angle by Kandlikar's model as follows

$$K = \frac{1 + \cos \theta}{16} \left[ \frac{2}{\pi} + \frac{\pi}{4} (1 + \cos \theta) \right]^{0.5} \tag{41}$$

where $\theta$ is the contact angle. As plotted in Fig. 11, the horizontal blue dashed line and red dashed line is the predicted value from Kandlikar's model with the contact angle of $\theta=55°$ and Zuber's model with $K=0.12$. The relative deviation is about 0.6% between the currently simulated CHF and the Zuber's model, which proves that the current hybrid phase-change model agrees well with Zuber's model for predicting CHF. And there is still 9.9% error between the current numerical model and the Kandlikar's model with $\theta=55°$. The much higher deviation is caused by the inadequate prediction of three-phase contact angle because, for the transient boiling process, it is rather difficult to evaluate a precise preceding contact angel due to the complicated interaction between the bubble and fluid [19]. However, the present results have a relatively high accuracy compared to previous research [19] carried out by the pseudopotential phase-change LBM model, which has a 16.6% error between the CHF and Zuber's model with $K=0.12$.

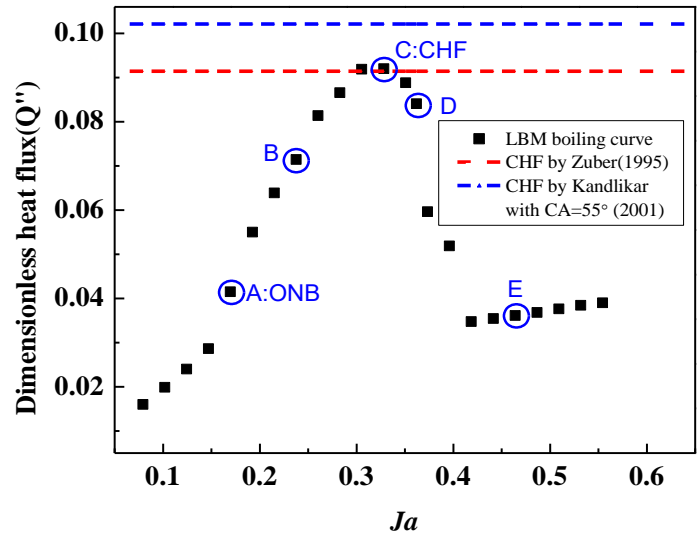

Fig. 11. Boiling curve obtained by the hybrid phase-change LBM model for a conjugated heat transfer under a constant temperature BC with nucleate spots in the heating surface ($\lambda_s/\lambda_L=50$, $\lambda_L/\lambda_v=17$, $\theta=55°$).

4.3.2. Different boiling regimes

In this section, to further investigate the different boiling regimes, parts of the entire boiling processes occurred in Fig. 11 are discussed in detail. Firstly, transient variation of the boiling process during the dimensionless time of $33.30<t^*<83.26$ corresponding to the point B in Fig. 11 are presented in Fig. 12 with a small wall superheat $Ja=0.2377$. One can observe from Fig. 12(a) that, a lot of isolated vapor slugs and columns grow in the heating surface. Subsequently, some of bubbles have departed from the heating surface due to the buoyancy force, and lots of small nucleate bubble grow on the top surface of the heater at the same time. Besides, parts of adjacent small bubbles coalesce together to yield larger bubble near the axis of $x=125$ and $x=275$ as shown in Fig. 12(c). The larger size of bubble near the axis of $x=270$ will depart from the top wall of heater because of the buoyancy force depicted in Fig. 12(d). Later, large number of bubbles have departed from the heating plate, which can be observed by Fig. 12(e). Furthermore, numerous small nucleate bubbles also generate on the top heating surface. Therefore, according to the simulated results in Fig. 12, it can be concluded that the boiling characteristic in current wall overheat is featured by the fully developed nucleate boiling pattern, which is also consistent with the boiling curve in Fig. 11.

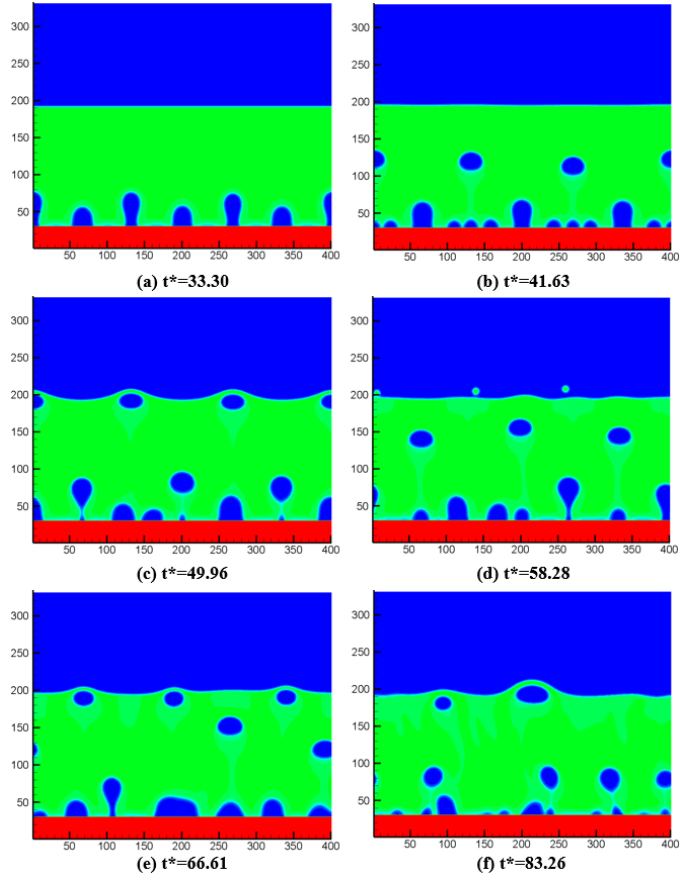

Fig. 12. Snapshots of the boiling characteristics with a low wall superheat (Point B, $Ja = 0.2377$).

Fig. 13 shows the transient variation of boiling process at the point of C with the bottom surface of the heater having wall superheat $Ja = 0.3282$, which also corresponds to the CHF point in Fig. 11. As shown in Fig. 13(a), with the increment of wall superheat, the heating wall is almost covered by the generated bubbles, and the isolated bubbles are quite larger than the bubbles in Fig. 12(a) at the same dimensionless time of $t^*=33.30$. In the next time, the rising bubbles in Fig. 12(a) have reached in the top surface of liquid domain, which indicates that the formation of nucleate bubble has a higher ratio compared to the Fig. 11(a) and (b). As the thermal energy releases, lots of bubbles grow and depart from the top wall of the heater, which can be observed in Fig. 13(c) and (d). These behaviors also demonstrate that there is higher boiling heat transfer compared to the nucleate boiling in Fig. 12. In the meanwhile, two thin vapor film also generate in the time of $t^*=69.94$ due to the coalescence of adjacent bubbles in the regions of $60 < x < 150$ and $250 < x < 330$. Subsequently, due to the bubbles departed from the heating surface, vapor film becomes thinner as shown in Fig. 13(f).

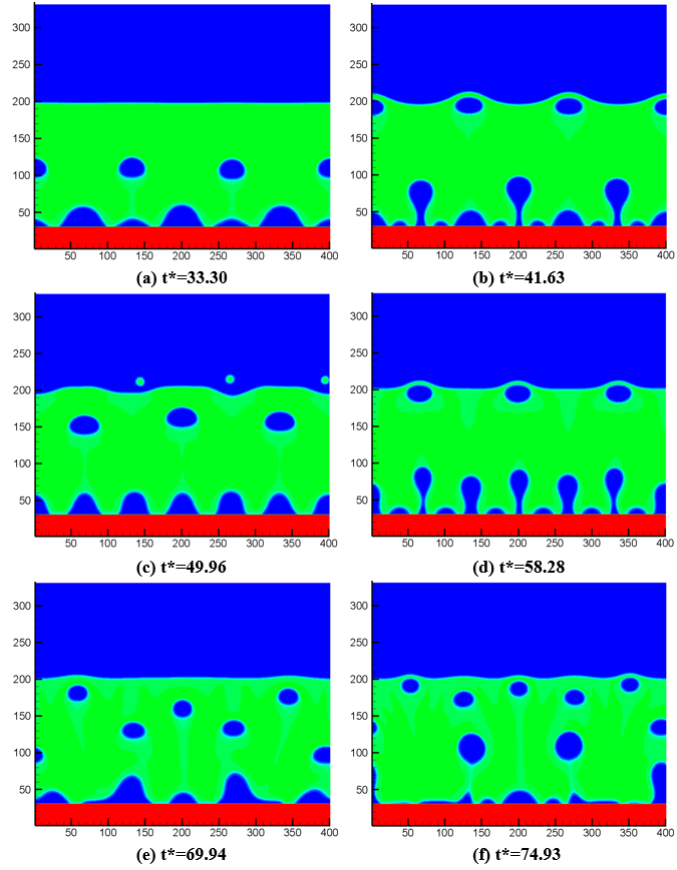

Fig. 13. Snapshots of the boiling process in the CHF regime with slightly high wall superheat (Point C, $Ja = 0.3282$).

The time evolutions of boiling process when the wall overheat is slightly larger than the point C corresponding to point D in Fig. 11 are clearly given by Fig. 14. At the initial stage, the density contour is almost the same as Fig. 13(a) at the time of $t^*=33.30$. However, due to the slight larger wall superheat, there is no small bubble formation and lots of thin vapor film yield in the heating surface as shown in Fig. 14(b-d), which suppresses the heat transfer from the solid to liquid. This boiling characteristic results in the decreasing of the heat flux dramatically due to the low thermal conductivity of thin vapor film. Therefore, it can be concluded that this boiling process has reached to the transition boiling regime. Furthermore, the transverse movement of the bubbles could also be observed in the axis of $x=50$ and $x=300$ as presented by Fig. 14 (e). This is due to the influence of natural convection and Marangoni effect, and this dynamic behavior agrees well with experimental results [69, 70]. At the same time. As shown in Fig. 14 (f), both nucleate bubbles and thin film vapor are yielding in the heating surface. Therefore, it could be inferred that the transition boiling regime can be regarded as the combination of the nucleate boiling and film boiling.

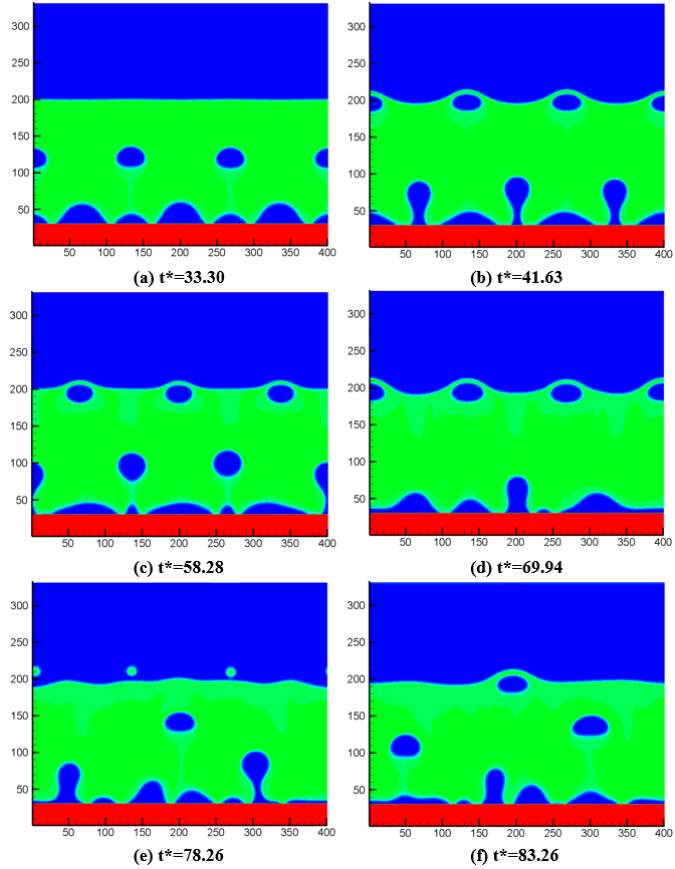
Fig. 14. Temporal evolutions of the transition boiling regime with a high wall superheat (Point D, $Ja = 0.3622$).

The transient variations of dimensionless heat flux for the different boiling regimes from the nucleate boiling (point B in Fig. 11) to CHF (point C in Fig. 11), to transient boiling (point D in Fig. 11) to film boiling (point E in Fig. 11) are plotted in Fig. 15 under the computational model B with the treatment of nucleate spots and constant temperature BC. One can observe in Fig. 15 that, there is a rather high fluctuation of heat flux in the CHF point and transition boiling pattern compared to the patterns of nucleate boiling and filming boiling, which fits well with the previous results conducted by LBM [19, 35]. This phenomenon is attributed to the partial dry out of the top wall in the heating plate and the unstable boiling process as demonstrated in Figs. (13) and (14).

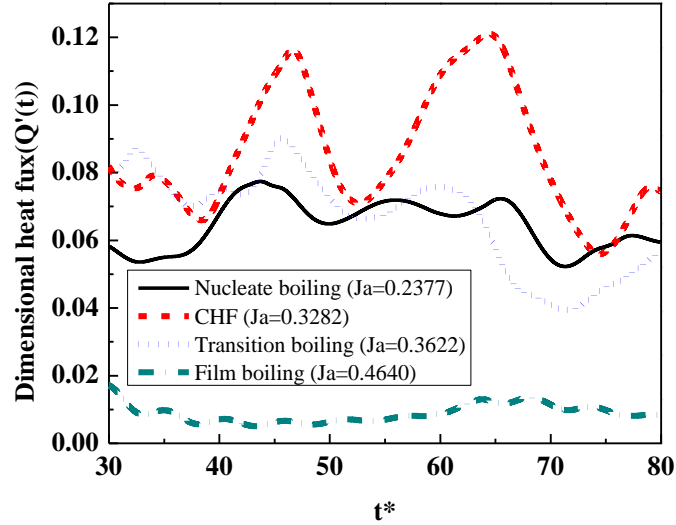

Fig. 15. Comparison of the transient dimensionless heat flux with time evolution for four different boiling patterns ( $\lambda_s / \lambda_L = 50$, $\lambda_L / \lambda_V = 17$, $\theta = 55°$ ).

*4.4. Thermal responses and heat transfer mechanism in different boiling regime*

Fig. 16 presents the transient temperature contour inside the heating substrate with a low wall overheat ( $Ja = 0.2377$ ) at the time of $t^*$=58.28 corresponding to Fig. 12(d). The local dimensionless temperature and corresponding local heat flux in the top surface of heating plate are also plotted in Fig. 17(a) and (b), respectively. Note that the dimensionless temperature is evaluated by $T^* = (T - T_{sat}) / (T_b - T_{sat})$. One can observe that, the isothermals beneath the small bubble have an no-uniformly distributed temperature contour and have a high penetration depth inside the solid domain of the heater, demonstrating that using a constant temperature in the bottom surface of fluid domain is not accurate. At the same time, the lower temperature regions in the top surface of the heater are located in the bottom surface of the small growing bubble as shown in Figs. 16 and 17(a), while slightly high-temperature areas appear beneath the liquid phase. In addition, as shown in Fig. 17(b), there are a lot of high heat fluxes in the regions of the triple-phase contact line as denoted by the blue circles. This is caused by the microlayer evaporation and triple-phase contact line evaporation in the nucleate boiling regime with a low wall overheat.

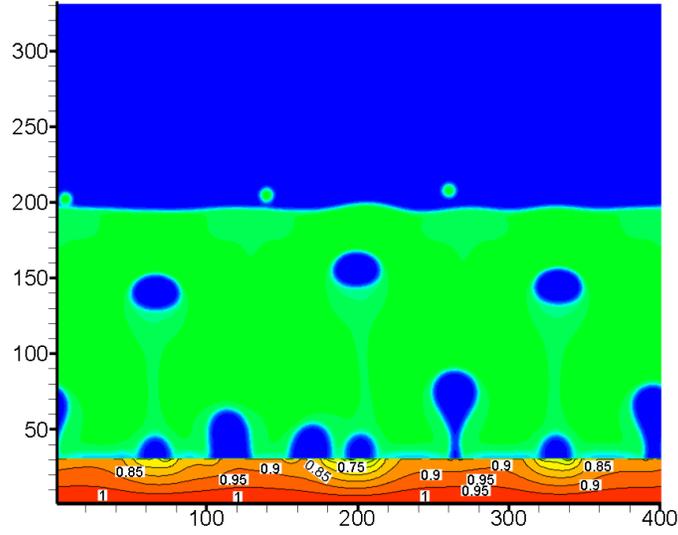

Fig. 16. Thermal contour inside the heater for the nucleate boiling regime at the dimensionless time of $t^*$=58.28 (Point B in Fig. 11, $Ja = 0.2377$ ).

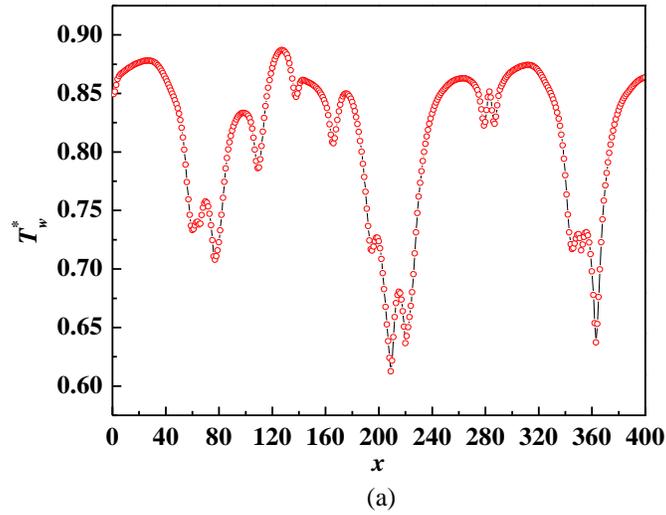

(a)

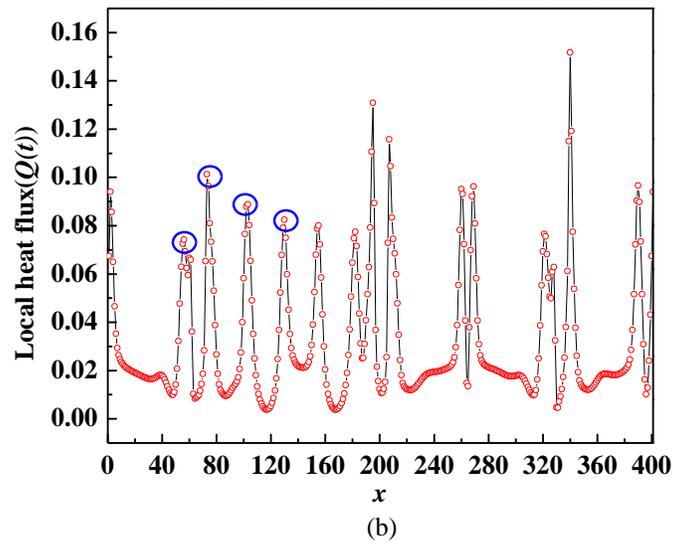

(b)

Fig. 17. Dimensional temperature variation (a) and local heat flux distribution (b) in the top surface of the heater for the nucleate boiling process (Point B from Fig. 11, $Ja = 0.2377$).

Fig. 18 illustrates the thermal responses inside the solid domain at the dimensionless time of $t^*=69.94$ with a slightly high wall overheat ($Ja = 0.3282$), which also corresponds to Fig. 13(e) and CHF in Fig. 11. The corresponding dimensionless temperature and local heat flux are given by Fig. 19(a) and (b), respectively. It can be observed from Fig. 18 that, the higher temperature areas in the top surface of the heater are mainly located in the bottom of bubbles, in which the heat flux is also lower as shown in Fig. 19(b) (blue dotted circles). Meantime, the lower temperature areas still appear in the triple-phase contact line with a high heat flux as shown in Fig. 19(b). It is also found that, there are highly fluctuant heat fluxes in the regions of nucleate spots when introducing an inconstant wettability, resulting in a unbalance interaction force.

Fig. 20 further presents the transient variation of temperature contour inside the heating plate with a high wall overheat ($Ja = 0.3622$) corresponding to Fig. 14(d). And the corresponding dimensionless temperature and local heat flux beneath fluid domain are also illustrated in Fig. 21(a) and (b), respectively. As shown in Fig. 20 that, the higher temperature areas are occupied a larger part of solid domain compared to Figs. 16 and 18. At the same time, the higher temperature regions in the top surface of the heater appear in the bottom of bubbles and thin vapor film unless the bottom surface of the small nucleate bubble having a slightly low temperature, which also can be seen in Fig. 21(a) (blue dotted circles). The lower heat flux is located in the region of bottom of thin vapor film and the center of the bubbles as presented in Fig. 21(b) (blue dotted circles). These behaviors are due to the transition boiling considered as the combination of nucleate boiling and film boiling, leading to the formation of the partial dryout and disappearance of the microlayer evaporation. Therefore, the low temperature areas inside the heater only appear in the triple-contact line in current boiling regime, which indicates that the phase-change appears in these regions. Accordingly, it could be concluded that the triple-contact line evaporation is the main boiling heat transfer mechanism in transition boiling regime.

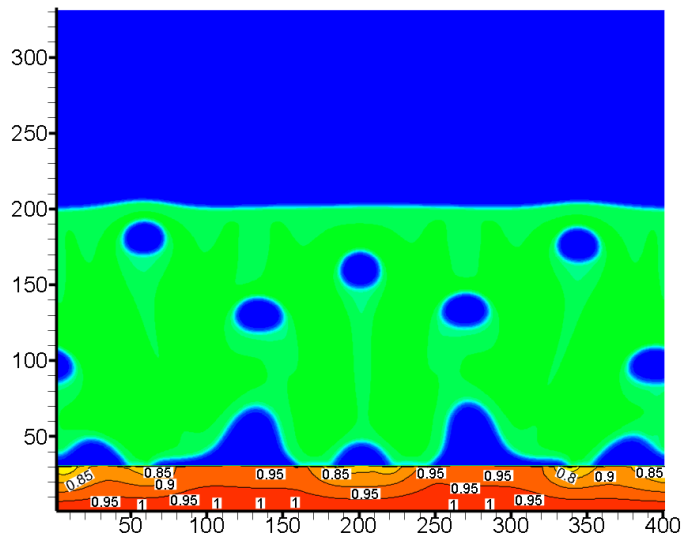

Fig. 18. Dimensional temperature contour inside the heating plate having a middle wall superheat (Point C in Fig. 11, $Ja = 0.3282$) at the dimensionless time of $t^*=69.94$.

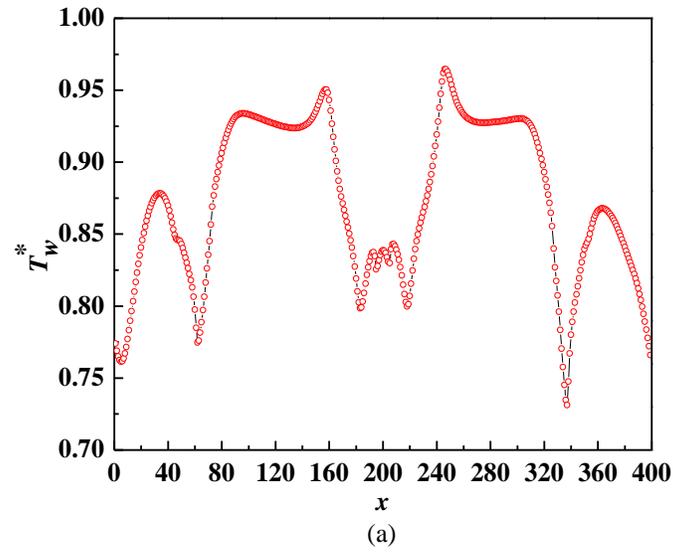

(a)

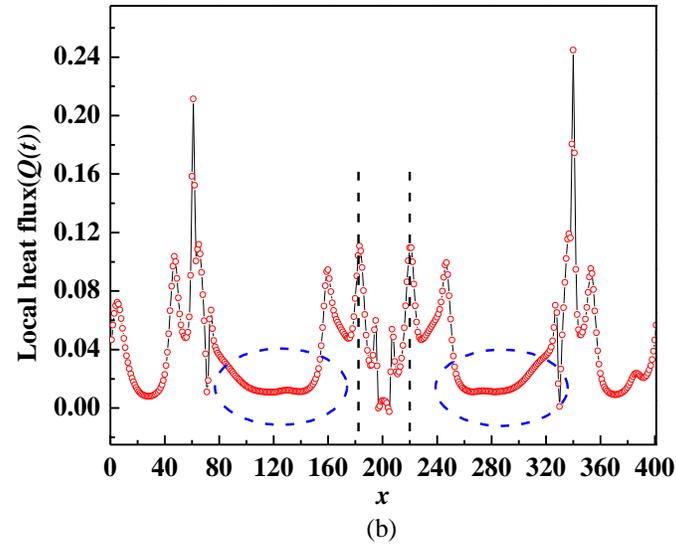

(b)

Fig. 19. Dimensional temperature variation (a) and local heat flux (b) in the top surface of the heater for the CHF (Point C in Fig. 11, $Ja = 0.3282$).

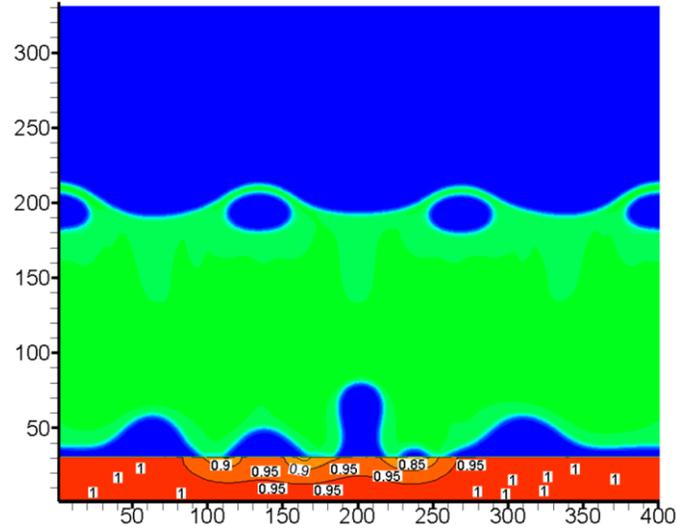

Fig. 20 Thermal response inside the heating substrate having a high wall superheat ($Ja = 0.3622$, Point D in Fig. 11) for the regime of transition boiling at the dimensionless time of $t^*=69.94$.

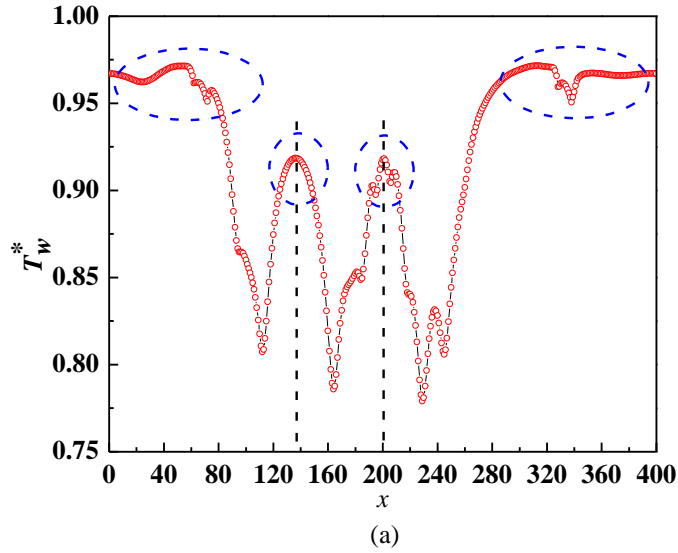

(a)

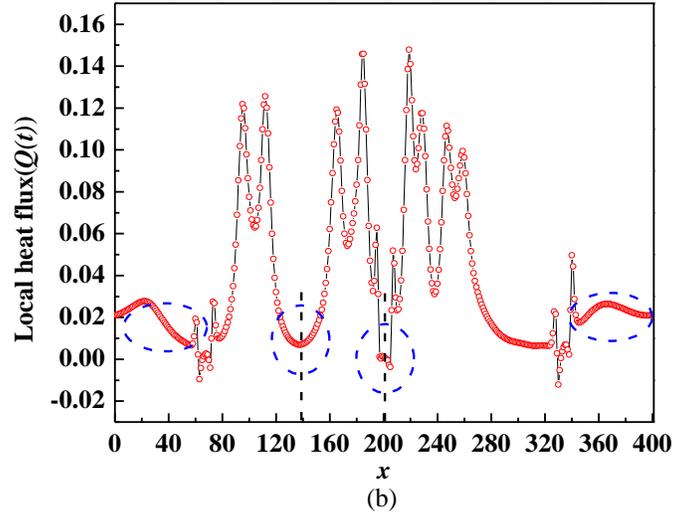

(b)

Fig. 21. Dimensional temperature distribution (a) and local heat flux (b) in the top surface of heating plate for the transition boiling process at the dimensionless time of $t^*=69.94$ ($Ja = 0.3622$, Point D in Fig. 11).

## 5. Conclusions

In this work, direct numerical simulations of the pool boiling heat transfer with and without considering conjugate heat transfer were investigated based on the improved pseudopotential hybrid phase-change. Effects of two treatments of the temperature BC in two computational models on pool boiling heat transfer have been studied in detail. The boiling curve from the ONB to fully developed film boiling have been obtained numerically. The temperature contour inside the heating substrate and local heat flux in the top surface of the heater were presented. The main conclusions of current investigations are given as follows:

1. Regarding the computational domain without conjugate heat transfer, effects of two treatments of temperature BC on the boiling characteristic of the nucleate boiling with a small degree of superheat are almost the same, whereas the treatments of constant temperature BC with nucleate spots is quite easier to produce a film boiling compared to the temperature BC with a small fluctuation at the same slightly high wall superheat.

2. With respect to the computational domain having conjugate heat transfer, the treatments of BC with constant temperature and nucleate spots are much easier to yield nucleate bubbles compared to the fluctuant temperature BC during the initial time step. Besides, the treatment of temperature BC with a fluctuant temperature is beneficial to generate film boiling heat transfer compared to the treatment of temperature BC with constant temperature and nucleate spots. This conclusion is the opposite for the case without considering conjugate heat transfer.

3. The entire boiling curve is numerically obtained by the improved pseudopotential hybrid phase-change LBM based on the computational domain considering conjugated heat transfer. The simulated CHF agrees well with the previous analytical solutions from the Zuber's model [10] and Kandlikar's model [9]. Hence, preciseness of the pseudopotential hybrid phase-change LBM is quantitively verified.

4. The transition boiling pattern can be regarded as the combination of the nucleate boiling and the film boiling. And the transient heat flux from the CHF boiling and transition boiling have a rather higher fluctuant characteristic compared to nucleate boiling and film boiling.

5. The low-temperature areas are located in the bottom surface of the bubble at a low wall overheat, whereas for the pooling boiling at a high wall superheat, the low temperature only appears in the three-triple contact line. High heat flux for pool boiling mainly comes from the triple-phase contact line evaporation, and the areas of the nucleate spots have a high fluctuant heat flux.

**Acknowledgments**

This work was supported by the National Natural Science Foundation of China (No.). The first author is grateful to the China Scholarship Council (CSC) for financially support (No.201806820023).

**References**

[1] V. Dhir, Boiling heat transfer, Annual Review of Fluid Mechanics, 30 (1998) 365-401.
[2] V.P. Carey, Liquid vapor phase change phenomena: an introduction to the thermophysics of vaporization and condensation processes in heat transfer equipment, CRC Press, 2018.
[3] S. Nukiyama, The maximum and minimum values of the heat Q transmitted from metal to boiling water under atmospheric pressure, International Journal of Heat and Mass Transfer, 9 (1966) 1419-1433.
[4] F.D. Moore, R.B. Mesler, The measurement of rapid surface temperature fluctuations during nucleate boiling of water, Aiche Journal, 7 (1961) 620-624.
[5] W.M. Rohsenow, A method of correlating heat transfer data for surface boiling of liquids, in, Cambridge, Mass.: MIT Division of Industrial Cooporation,[1951], 1951.
[6] J. Jung, S.J. Kim, J. Kim, Observations of the critical heat flux process during pool boiling of FC-72, Journal of Heat Transfer, 136 (2014) 041501.
[7] D.E. Kim, J. Song, H. Kim, Simultaneous observation of dynamics and thermal evolution of irreversible dry spot at critical heat flux in pool boiling, International Journal of Heat and Mass Transfer, 99 (2016) 409-424.
[8] X. Quan, G. Chen, P. Cheng, A thermodynamic analysis for heterogeneous boiling nucleation on a superheated wall, International Journal of Heat and Mass Transfer, 54 (2011) 4762-4769.
[9] S.G. Kandlikar, A theoretical model to predict pool boiling CHF incorporating effects of contact angle and orientation, Journal of Heat Transfer, 123 (2001) 1071-1079.
[10] N. Zuber, Hydrodynamic aspects of boiling heat transfer (thesis), in, Ramo-Wooldridge Corp., Los Angeles, CA (United States); Univ. of California …, 1959.

[11] Y. Haramura, Y. Katto, A new hydrodynamic model of critical heat flux, applicable widely to both pool and forced convection boiling on submerged bodies in saturated liquids, International Journal of Heat and Mass Transfer, 26 (1983) 389-399.
[12] P.J. Berenson, Film-Boiling Heat Transfer From a Horizontal Surface, Journal of Heat Transfer, 83 (1961) 351-356.
[13] G. Son, V.K. Dhir, N. Ramanujapu, Dynamics and heat transfer associated with a single bubble during nucleate boiling on a horizontal surface, Journal of Heat Transfer, 121 (1999) 623-631.
[14] A. Mukherjee, V. Dhir, Study of lateral merger of vapor bubbles during nucleate pool boiling, Transactions of the ASME-C-Journal of Heat Transfer, 126 (2004) 1023-1039.
[15] D. Juric, G. Tryggvason, Computations of boiling flows, International Journal of Multiphase Flow, 24 (1998) 387-410.
[16] S.W. Welch, J. Wilson, A volume of fluid based method for fluid flows with phase change, Journal of Computational Physics, 160 (2000) 662-682.
[17] J. Kim, Review of nucleate pool boiling bubble heat transfer mechanisms, International Journal of Multiphase Flow, 35 (2009) 1067-1076.
[18] V.K. Dhir, G.R. Warrier, E. Aktinol, Numerical simulation of pool boiling: a review, Journal of Heat Transfer, 135 (2013) 061502.
[19] S. Gong, P. Cheng, Direct numerical simulations of pool boiling curves including heater's thermal responses and the effect of vapor phase's thermal conductivity, International Communications in Heat & Mass Transfer, 87 (2017) 61-71.
[20] L. Chen, Q. Kang, Y. Mu, Y.L. He, W.Q. Tao, A critical review of the pseudopotential multiphase lattice Boltzmann model: Methods and applications, International Journal of Heat & Mass Transfer, 76 (2014) 210-236.
[21] X. Shan, H. Chen, Lattice Boltzmann model for simulating flows with multiple phases and components, Phys Rev E Stat Phys Plasmas Fluids Relat Interdiscip Topics, 47 (1993) 1815-1819.
[22] Q. Li, K.H. Luo, Q.J. Kang, Y.L. He, Q. Chen, Q. Liu, Lattice Boltzmann methods for multiphase flow and phase-change heat transfer, Progress in Energy & Combustion Science, 52 (2016) 62-105.
[23] S. Gong, P. Cheng, A lattice Boltzmann method for simulation of liquid–vapor phase-change heat transfer, International Journal of Heat and Mass Transfer, 55 (2012) 4923-4927.
[24] S. Gong, P. Cheng, Lattice Boltzmann simulation of periodic bubble nucleation, growth and departure from a heated surface in pool boiling, International Journal of Heat & Mass Transfer, 64 (2013) 122-132.
[25] S. Gong, P. Cheng, Lattice Boltzmann simulations for surface wettability effects in saturated pool boiling heat transfer, International Journal of Heat and Mass Transfer, 85 (2015) 635-646.
[26] C. Zhang, P. Cheng, Mesoscale simulations of boiling curves and boiling hysteresis under constant wall temperature and constant heat flux conditions, International Journal of Heat and Mass Transfer, 110 (2017) 319-329.
[27] X. Ma, P. Cheng, 3D simulations of pool boiling above smooth horizontal heated surfaces by a phase-change lattice Boltzmann method, International Journal of Heat and Mass Transfer, 131 (2019) 1095-1108.
[28] X. Ma, P. Cheng, X. Quan, Simulations of saturated boiling heat transfer on bio-inspired two-phase heat sinks by a phase-change lattice Boltzmann method, International Journal of Heat and Mass Transfer, 127 (2018) 1013-1024.
[29] X. Liu, P. Cheng, Lattice Boltzmann simulation for dropwise condensation of vapor along vertical hydrophobic flat plates, International Journal of Heat & Mass Transfer, 64 (2013) 1041-1052.
[30] X. Liu, C. Ping, Dropwise condensation theory revisited: Part I. Droplet nucleation radius, International Journal of Heat & Mass Transfer, 83 (2015) 833-841.
[31] X. Li, P. Cheng, Lattice Boltzmann simulations for transition from dropwise to filmwise condensation on hydrophobic surfaces with hydrophilic spots, International Journal of Heat and Mass Transfer, 110 (2017) 710-722.
[32] C. Zhang, F. Hong, P. Cheng, Simulation of liquid thin film evaporation and boiling on a heated hydrophilic microstructured surface by Lattice Boltzmann method, International Journal of Heat & Mass Transfer, 86 (2015) 629-638.
[33] W.-Z. Fang, L. Chen, Q.-J. Kang, W.-Q. Tao, Lattice Boltzmann modeling of pool boiling with large liquid-gas density ratio, International Journal of Thermal Sciences, 114 (2017) 172-183.
[34] Y.-T. Mu, L. Chen, Y.-L. He, Q.-J. Kang, W.-Q. Tao, Nucleate boiling performance evaluation of cavities at mesoscale level, International Journal of Heat and Mass Transfer, 106 (2017) 708-719.
[35] Q. Li, Q.J. Kang, M.M. Francois, Y.L. He, K.H. Luo, Lattice Boltzmann modeling of boiling heat transfer: The boiling curve and the effects of wettability, International Journal of Heat and Mass Transfer, 85 (2015) 787-796.
[36] Q. Li, P. Zhou, H. Yan, Pinning-depinning mechanism of the contact line during evaporation on chemically patterned surfaces: A lattice Boltzmann study, Langmuir the Acs Journal of Surfaces & Colloids, 32 (2016) 9389.

[37] Q. Li, Y. Yu, P. Zhou, H. Yan, Enhancement of boiling heat transfer using hydrophilic-hydrophobic mixed surfaces: A lattice Boltzmann study, Applied Thermal Engineering, (2017).
[38] Y. Yu, Q. Li, C.Q. Zhou, P. Zhou, H. Yan, Investigation of droplet evaporation on heterogeneous surfaces using a three-dimensional thermal multiphase lattice Boltzmann model, Applied Thermal Engineering, 127 (2017) 1346-1354.
[39] Y. Yu, Z. Wen, Q. Li, P. Zhou, H. Yan, Boiling heat transfer on hydrophilic-hydrophobic mixed surfaces: A 3D lattice Boltzmann study, Applied Thermal Engineering, 142 (2018) 846-854.
[40] X. Shan, H. Chen, Simulation of nonideal gases and liquid-gas phase transitions by the lattice Boltzmann equation, Phys Rev E Stat Phys Plasmas Fluids Relat Interdiscip Topics, 49 (1994) 2941-2948.
[41] H.N.W. Lekkerkerker, C.K. Poon, P.N. Pusey, A. Stroobants, P.B. Warren, Lattice BGK Models for Navier-Stokes Equation, Europhysics Letters, (1996) 479-484.
[42] P. Lallemand, L.-S. Luo, Theory of the lattice Boltzmann method: Dispersion, dissipation, isotropy, Galilean invariance, and stability, Physical Review E, 61 (2000) 6546-6562.
[43] Q. Li, K.H. Luo, X.J. Li, Lattice Boltzmann modeling of multiphase flows at large density ratio with an improved pseudopotential model, Phys Rev E Stat Nonlin Soft Matter Phys, 87 (2013) 053301.
[44] M.E. Mccracken, J. Abraham, Multiple-relaxation-time lattice-Boltzmann model for multiphase flow, Physical Review E Statistical Nonlinear & Soft Matter Physics, 71 (2005) 036701.
[45] Q. Li, Y.L. He, G.H. Tang, W.Q. Tao, Improved axisymmetric lattice Boltzmann scheme, Physical Review E Statistical Nonlinear & Soft Matter Physics, 81 (2010) 056707.
[46] A. Xu, T.S. Zhao, L. An, L. Shi, A three-dimensional pseudo-potential-based lattice Boltzmann model for multiphase flows with large density ratio and variable surface tension, International Journal of Heat & Fluid Flow, 56 (2015) 261-271.
[47] A. Xu, T.S. Zhao, L. Shi, X.H. Yan, Three-dimensional lattice Boltzmann simulation of suspensions containing both micro- and nanoparticles, International Journal of Heat & Fluid Flow, (2016).
[48] W. Zhao, Y. Zhang, B. Xu, P. Li, Z. Wang, S. Jiang, Multiple-Relaxation-Time Lattice Boltzmann Simulation of Flow and Heat Transfer in Porous Volumetric Solar Receivers, Journal of Energy Resources Technology, 140 (2018) 082003.
[49] Q. Liu, Y.L. He, Q. Li, W.Q. Tao, A multiple-relaxation-time lattice Boltzmann model for convection heat transfer in porous media, International Journal of Heat & Mass Transfer, 73 (2014) 761-775.
[50] Q. Li, K.H. Luo, Achieving tunable surface tension in the pseudopotential lattice Boltzmann modeling of multiphase flows, Physical Review E Statistical Nonlinear & Soft Matter Physics, 88 (2013) 053307.
[51] W. Zhao, Y. Zhang, W. Shang, Z. Wang, B. Xu, S. Jiang, Simulation of droplet impacting a square solid obstacle in microchannel with different wettability by using high density ratio pseudopotential multiple-relaxation-time (MRT) lattice Boltzmann method (LBM), Canadian Journal of Physics, 97 (2019) 93-113.
[52] W. Zhao, Y. Zhang, B. Xu, An improved pseudopotential multi-relaxation-time lattice Boltzmann model for binary droplet collision with large density ratio, Fluid Dynamics Research, 51 (2019) 025510.
[53] P. Lallemand, L.S. Luo, Theory of the lattice boltzmann method: dispersion, dissipation, isotropy, galilean invariance, and stability, Physical Review E Statistical Physics Plasmas Fluids & Related Interdisciplinary Topics, 61 (2000) 6546-6562.
[54] Q. Li, K.H. Luo, Thermodynamic consistency of the pseudopotential lattice Boltzmann model for simulating liquid–vapor flows, Applied Thermal Engineering, 72 (2014) 56-61.
[55] P. Yuan, L. Schaefer, Equations of state in a lattice Boltzmann model, Physics of Fluids, 18 (2006) 329.
[56] Q. Li, K.H. Luo, Q.J. Kang, Q. Chen, Contact angles in the pseudopotential lattice Boltzmann modeling of wetting, Phys.rev.e, 90 (2014) 053301.
[57] A. Hu, D. Liu, Study of boiling heat transfer on the wall with an improved hybrid lattice Boltzmann model, arXiv preprint arXiv:1812.10375, (2018).
[58] R. Zhang, H. Chen, Lattice Boltzmann method for simulations of liquid-vapor thermal flows, Physical Review E, 67 (2003) 066711.
[59] Q. Li, P. Zhou, H. Yan, Improved thermal lattice Boltzmann model for simulation of liquid-vapor phase change, Physical Review E, 96 (2017) 063303.
[60] H. Liu, A.J. Valocchi, Y. Zhang, Q. Kang, Phase-field-based lattice Boltzmann finite-difference model for simulating thermocapillary flows, Phys Rev E Stat Nonlin Soft Matter Phys, 87 (2013) 013010.
[61] T. Lee, C.-L. Lin, A stable discretization of the lattice Boltzmann equation for simulation of incompressible two-phase flows at high density ratio, Journal of Computational Physics, 206 (2005) 16-47.

[62] Y. Yu, Q. Li, C.Q. Zhou, P. Zhou, H.J. Yan, Y. Yu, Q. Li, C.Q. Zhou, P. Zhou, H.J. Yan, Investigation of droplet evaporation on heterogeneous surfaces using a three-dimensional thermal multiphase lattice Boltzmann model, Applied Thermal Engineering, 127 (2017).

[63] W. Gong, Y. Yan, S. Chen, E. Wright, A modified phase change pseudopotential lattice Boltzmann model, International Journal of Heat and Mass Transfer, 125 (2018) 323-329.

[64] Q. Li, P. Zhou, H.J. Yan, Improved thermal lattice Boltzmann model for simulation of liquid-vapor phase change, (2017).

[65] A. Hu, D. Liu, A superheat degree driven liquid-vapor phase-change lattice Boltzmann model, International Journal of Heat and Mass Transfer, 136 (2019) 674-680.

[66] X. He, Q. Zou, L.-S. Luo, M. Dembo, Analytic solutions of simple flows and analysis of nonslip boundary conditions for the lattice Boltzmann BGK model, Journal of Statistical Physics, 87 (1997) 115-136.

[67] G. Házi, A. Márkus, Modeling heat transfer in supercritical fluid using the lattice Boltzmann method, Physical Review E, 77 (2008) 026305.

[68] X. Chang, H. Huang, Y.-P. Cheng, X.-Y. Lu, Lattice Boltzmann study of pool boiling heat transfer enhancement on structured surfaces, 2019.

[69] H. Wang, X. Peng, B. Wang, D. Lee, Jet flow phenomena during nucleate boiling, International Journal of Heat and Mass Transfer, 45 (2002) 1359-1363.

[70] H. Wang, X. Peng, B. Wang, D. Lee, Bubble sweeping and jet flows during nucleate boiling of subcooled liquids, International Journal of Heat and Mass Transfer, 46 (2003) 863-869.